\documentclass[lettersize,journal]{IEEEtran}
\usepackage{amsmath,amsfonts}
\usepackage{algorithmic}
\usepackage{algorithm}
\usepackage{array}
\usepackage[caption=false,font=normalsize,labelfont=sf,textfont=sf]{subfig}
\usepackage{textcomp}
\usepackage{stfloats}
\usepackage{url}
\usepackage{verbatim}
\usepackage{graphicx}
\usepackage{cite}
\usepackage{amsfonts}
\usepackage[numbers,sort&compress]{natbib}
\makeatletter
\renewcommand{\maketag@@@}[1]{\hbox{\m@th\normalsize\normalfont#1}}%
\makeatother
\begin{document}

\title{Hybrid Precoding with Per-Beam Timing Advance for Asynchronous Cell-free mmWave Massive MIMO-OFDM Systems}

\author{Pengzhe~Xin,
		Yang~Cao,
		Yue~Wu,
		Dongming~Wang,~\IEEEmembership{Member,~IEEE,}
		Xiaohu~You,~\IEEEmembership{Fellow,~IEEE}
		and~Jiangzhou~Wang,~\IEEEmembership{Fellow,~IEEE}
\thanks{This work was supported by the Fundamental Research Funds for the Central Universities under Grant 2242022k60006.\emph{(Corresponding Author: Dongming Wang)}}
\thanks{P. Xin, Y. Cao, D. Wang, X. You and J. Wang are with National Mobile Communication Research Laboratory, Southeast University, Nanjing, 210096, China (email:\{xinpz,epstwxv,wangdm,xhyu,j.z.wang\}@seu.edu.cn).}
\thanks{Y. Cao, Y. Wu, D. Wang, X. You and J. Wang are also with Purple Mountain Laboratories, Nanjing, 210096, China.}
}



\maketitle

\begin{abstract}
Cell-free massive multiple-input-multiple-output (CF-mMIMO) is regarded as one of the promising technologies for next-generation wireless networks. However, due to its distributed architecture, geographically separated access points (APs) jointly serve a large number of user-equipments (UEs), there will inevitably be a discrepancies in the arrival time of transmitted signals. In this paper, we investigate millimeter-wave (mmWave) CF-mMIMO orthogonal frequency division multiplexing (OFDM) systems with asynchronous reception in a wide area coverage scenario, where asynchronous timing offsets may extend far beyond the cyclic prefix (CP) range. A comprehensive asynchronous beam-domain signal transmission model is presented for mmWave CF-mMIMO-OFDM systems in both downlink and uplink, incorporating phase offset, inter-carrier interference (ICI) and inter-symbol interference (ISI). To address the issue of asynchronous reception, we propose a novel per-beam timing advance (PBTA) hybrid precoding architecture and analyze the spectral efficiency (SE) in the beam domain for downlink and uplink asynchronous receptions. Both scalable centralized and distributed implementations are taken into account, and the asynchronous delay phase is utilized to design precoding/combining vectors. Furthermore, we formulate the sum rate maximization problem and develop two low-complexity joint beam selection and UE association algorithms for the asynchronous scenario, which take into account the impact of asynchronous timing offset exceeding the CP range. Simulation results demonstrate that the performance will be severely limited by ICI and ISI, and our proposed PBTA hybrid precoding architecture effectively mitigates asynchronous interference compared to the nearest AAU/UE-based timing-advance scheme. Additionally, numerical results show that our proposed low-complexity joint beam selection and UE association algorithms achieve superior SE performance.
\end{abstract}


\section{Introduction}
\IEEEPARstart{W}{ith} the ongoing advancement of communication technology and the escalating demand for data transmission, the fifth-generation (5G) mobile communication system has been extensively promoted. Both academia and industry are also embarking on the exploration of the sixth-generation (6G) technology to meet the higher data communication needs in the future \cite{you2021towards,ammar2021user}. Cell-free massive multiple-input-multiple-output (CF-mMIMO) is regarded as an important candidate technology for 6G communication networks, surpassing the constraints of conventional cell boundaries. It holds the potential to extend network coverage beyond that of cellular MIMO, offering enhanced quality of service for users \cite{elhoushy2021cell,liu2019spectral}. In contrast to traditional mMIMO and distributed antenna systems at the same location, the fundamental concept of CF-mMIMO systems lies in multiple geographically dispersed access points (APs) linked to a central processing unit (CPU) consistently providing services to user equipments (UEs) through spatial multiplexing across the same time-frequency resources \cite{ngo2017cell,chen2020structured}. All APs are connected to the CPU via fronthaul links, with the central processor tasked with coordinating and processing data between UEs and APs \cite{bjornson2020scalable}. By leveraging macro diversity gains and employing interference mitigation techniques in cellular mMIMO systems, the CF-mMIMO wireless network is poised to attain superior spectral efficiency (SE) and energy efficiency to traditional small-cell and cellular wireless networks \cite{CHEN2022695,cao2023experimental}.

CF-mMIMO system is a distributed network architecture where distributed processing is essential. Two network operation methods with different levels of cooperation have been proposed for data processing, namely centralized operation and distributed operation \cite{bjornson2019making}. In centralized operation, channel estimation and data detection are jointly executed at the CPU, whereas in distributed operation, all tasks except the final data detection are conducted at the APs. The authors in \cite{bjornson2019making} compared various uplink data detection methods for CF-mMIMO, including centralized and distributed processing, and demonstrated that under centralized operation, the uplink SE significantly surpasses that achieved under distributed operation. \cite{bjornson2020scalable} proposed a scalable partial serving strategy based on a user-centered dynamic cooperation clustering (DCC) association method. This approach offers a novel algorithm for joint initial access, pilot allocation and cluster formation. Moreover, \cite{wang2023full} proposed a scalable architecture for CF-mMIMO implementation, utilizing distributed transceivers and scalable cooperative transmission.

By tapping into the underutilized bandwidth within the millimeter-wave (mmWave) spectrum, mmWave communications can attain multi-Gbps data transmission rates \cite{heath2016overview}. Thus, the integration of mmWave communications with cell-free networks presents a promising pathway for the next-generation wireless communications, enabling ultra-high peak rates while enhancing SE \cite{alonzo2017cell,alonzo2019energy,femenias2019cell,meyer2022state,yuan20223d}. MmWave frequency bands not only offer wider available bandwidths but also enable the placement of more antennas within a small area, owing to half-wavelength spacing \cite{xiao2017millimeter}. Furthermore, the significant path loss associated with high-frequency transmission over long distances can be mitigated by employing highly directional beamforming with large-scale antenna arrays \cite{dai2022delay,rappaport2013millimeter,liu2014phase}. Hybrid precoding schemes in mmWave massive MIMO systems is a promising research direction \cite{nguyen2022hybrid,gao2016energy,sohrabi2016hybrid}. Hybrid precoding is divided into digital precoding, which requires a reduced number of RF chains, and extensive analog precoding, achieved through analog phase shifters. Additionally, a particularly promising method for implementing analog precoding involves using a lens antenna array, which consists of a lens and an antenna array positioned on the focal surface of the lens \cite{brady2013beamspace,amadori2015low,guo2018joint}. Utilizing the discrete lens array (DLA), which incurs negligible performance loss \cite{zeng2016millimeter}, beamspace MIMO can convert the traditional spatial channel into the beamspace channel, effectively capturing channel sparsity at mmWave frequencies.

Most current studies on CF-mMIMO focus on synchronous scenarios, assuming that each receiver can simultaneously receive signals from all transmitters. However, this assumption is impractical in a distributed architecture. Given the different locations of the APs and UEs, there will inevitably be differences in the arrival time of the transmitted signals. Therefore, interference in CF-mMIMO is inherently asynchronous. Especially in the scenario of wide area coverage and ubiquitous connection, the asynchronous timing offset may cause the arrival time of signals to exceed the cyclic prefix (CP) range. This results in severe phase shift, inter-carrier interference (ICI) and inter-symbol interference (ISI), which seriously degrade the performance of the CF-mMIMO orthogonal frequency division multiplexing (OFDM) \cite{zhu2009chunk,zhu2011chunk} system. \cite{yan2019asynchronous} analyzed the impacts of asynchronous reception on distributed massive MIMO-OFDM systems. \cite{li2021impacts} analyzed how phase shifts, resulting from differences in signal arrival times, affect the performance of channel estimation and downlink SE. The study utilized mutually orthogonal time-multiplexing pilot sequences for channel estimation while disregarding pilot contamination. \cite{li2023cell} developed a minimum mean square error (MMSE) channel estimation method for asynchronous reception and derived closed-form expressions for the uplink achievable rate using maximum ratio (MR) combining. \cite{zheng2023asynchronous} and \cite{zheng2022performance} investigated the downlink performance under asynchronous reception caused by both delay and oscillator phase variations, and apply the rate-splitting strategy to CF-mMIMO systems to compensate for the performance loss. Their findings indicate that asynchronous reception disrupts coherent data transmission, resulting in degraded SE. \cite{jafri2024cooperative} proposed a hybrid precoding scheme for asynchronous CF-mMIMO systems in mmWave scenarios to mitigate asynchronous interference.

However, only a few studies on CF-mMIMO-OFDM asynchronous interference, such as \cite{yan2019asynchronous}, \cite{li2023cell} and \cite{li2024cell}, have focused solely on performance analysis. These studies assumed that the likelihood of asynchronous timing offsets exceeding the CP range is minimal, thereby underestimating the impact of inter-carrier interference (ICI) and inter-symbol interference (ISI) on system performance. Moreover, the studies that attempted to mitigate and address the problem of asynchronous interference, such as \cite{li2021impacts}, \cite{zheng2023asynchronous}, \cite{zheng2022performance}, and \cite{jafri2024cooperative}, did not consider the OFDM scenario, focusing solely on the asynchronous phase shift, thereby ignoring the impact of ICI and ISI. Consequently, there is lack of performance analysis for asynchronous scenarios involving wide area coverage, where asynchronous timing offsets significantly exceed the CP range. Additionally, there is no effective solution for addressing asynchronous CF-mMIMO-OFDM scenarios that involve ICI and ISI.

Therefore, in view of the aforementioned problems, we propose a novel per-beam timing advance (PBTA) hybrid precoding architecture for asynchronous mmWave CF-mMIMO-OFDM scenarios. Specifically, APs are replaced by active antenna units (AAUs) in the CF-mMIMO system. Each AAU is equipped with a large-scale antenna array and can utilize a unified unitary matrix to transform spatial-domain signals into the beam domain. Given the sparsity of the mmWave channel and the spatial angle resolution of each UE, UEs located in different orientations can be distinguished. This enables multi-user spatial division transmission within the beam domain. Furthermore, inspired by \cite{you2017bdma}, our PBTA architecture compensates for path propagation delays, ensuring that signals reach each receiver approximately synchronously.

So far, no existing papers have attempted to leverage the characteristics of the beam domain to address the issue of asynchronous reception in CF-mMIMO-OFDM scenarios and achieve approximate synchronous reception.

The main contributions of this paper are summarized as follows:
\begin{itemize}
	\item First, we provide a comprehensive asynchronous beam-domain signal transmission model for mmWave CF-mMIMO-OFDM systems with hybrid precoding architecture for both downlink and uplink. MmWave frequency selective fading channel is adopted. Then, we analyze the impact of asynchronous reception on the CF-mMIMO-OFDM system under wide area coverage, focusing primarily on phase shift, ICI and ISI.
	\item Next, we propose a PBTA hybrid precoding architecture for asynchronous mmWave CF-mMIMO-OFDM scenarios, applicable to both downlink and uplink. In the proposed architecture, each AAU distinguishes different UEs in the beam domain and assigns corresponding beams. Each AAU then calculates the timing advance (TA) for different beams, subsequently performing PBTA to compensate for the asynchronous timing offset. This ensures that the signal is approximately synchronous upon reaching the receiver.
	\item We derive the achievable rates for both centralized and distributed implementations across four different scenarios: synchronization, asynchronization, PBTA and small cell, for both downlink and uplink. Furthermore, phase shift, ICI and ISI are incorporated into the expressions of achievable rates.
	\item Subsequently, we investigate the beam selection and UE association problem to maximize the ergodic achievable SE. Then we develop two suboptimal, low-complexity joint beam selection and UE association algorithms, which take into account the impact of asynchronous timing offset outside the CP range, mainly based on beam-domain channel amplitude and large-scale fading coefficient, respectively.
	\item Finally, numerical results are presented to validate the analytical results. We adopt the delay phase used MMSE and MR precoding and combining for downlink and uplink respectively. Both centralized and distributed implementations are taken into account. Extensive simulations verify that our proposed PBTA scheme significantly mitigates the interference caused by asynchronous reception. Additionally, it is also shown the superior performance of proposed joint beam selection and UE association algorithms.
\end{itemize}

The rest of this paper is organized as follows. In Section \ref{System and channel models}, we investigate the system model for mmWave CF-mMIMO-OFDM with hybrid precoding architecture, and the mmWave frequency selective fading channel model is introduced. In Section \ref{Downlink asynchronous reception}, we first propose PBTA hybrid precoding architecture for downlink asynchronous transmission. In Section \ref{Uplink Asynchronous reception}, we apply PBTA scheme to uplink asynchronous reception. In Section \ref{Beam selection and UE association}, two low-complexity joint beam selection and UE association algorithms are developed. Finally, simulation results are presented in In Section \ref{Simulation Results}, and the paper is concluded in Section \ref{Conclusion}.

\textit{Notation}: Uppercase and lowercase boldface letters are used to denote matrices and vectors, respectively. An $M \times M$ identity matrix is denoted by ${\bf{I}}_M$. The notations ${\mathbb{C}^{N \times M}}$ refer to complex ${N \times M}$ matrices. $\left|  \cdot  \right|$ denotes the absolute value of a scalar. ${\left[  \cdot  \right]^{\text{T}}}$, ${\left[  \cdot  \right]^{*}}$ and ${\left[  \cdot  \right]^{\text{H}}}$ represent the transpose, conjugate and Hermitian transpose of a vector or a matrix, respectively. ${\rm diag}({\bf x})$ is a diagonal matrix with $\bf x$ on its diagonal. We use ${\rm{blkdiag}}\left( {{{\bf{A}}_1}, \ldots ,{{\bf{A}}_n}} \right)$ for a block-diagonal matrix with the matrices ${{{\bf{A}}_1}, \ldots ,{{\bf{A}}_n}}$ on the diagonal. $\mathbb{E}\left[  \cdot  \right]$ represents mathematical expectation. The distribution of a circularly symmetric complex Gaussian random variable with zero mean and variance ${\sigma ^2}$ is denoted as ${\cal C}{\cal N}\left( {0,{\sigma ^2}} \right)$. We use $\left| {{\cal A}} \right|$ and $ \buildrel \Delta \over = $ to denote the cardinality of the set ${\cal A}$ and definitions respectively.

\section{System and channel models} \label{System and channel models}

\subsection{System Model}
\begin{figure}[!t]
	\centering
	\includegraphics[scale=0.55]{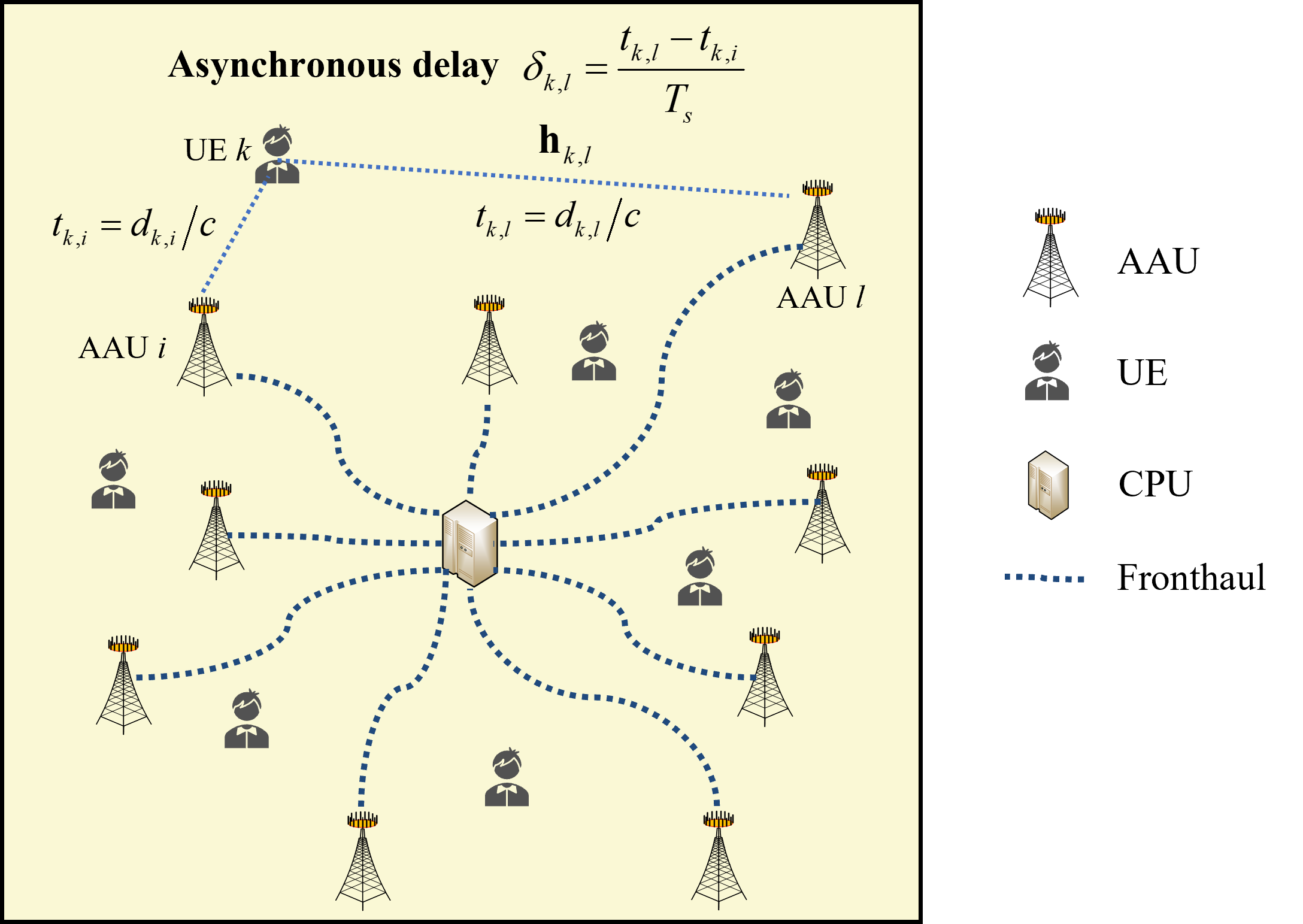}
	\caption{Asynchronous reception in a CF-mMIMO-OFDM system.}
	\label{figure1}
\end{figure}

Consider a mmWave CF-mMIMO-OFDM system with $M$ subcarriers as illustrated in Fig. \ref{figure1}, where $L$ AAUs and $K$ single-antenna UEs are randomly distributed in a large area. Each AAU is equipped with $N$ antennas and ${N_{{\rm{RF}}}}$ radio frequency (RF) chains, where ${N_{{\rm{RF}}}} \le N$ and ${N_{{\rm{RF}}}} \le K \le L{N_{{\rm{RF}}}}$. The sample length of the cyclic prefix is ${M_{{\text{CP}}}}$. We let ${\bf{s}} = {\left[ {{s_1}, \cdots ,{s_K}} \right]^{\rm{T}}}$ denote signals transmitted to $K$ UEs, where ${s_k}$ is the complex data symbol intended for UE $k$ with zero mean and normalized power of ${\mathbb{E}}\left\{ {{{\left| {{s_k}} \right|}^2}} \right\} = 1$. 

\begin{figure}[!t]
	\centering
	\includegraphics[scale=0.45]{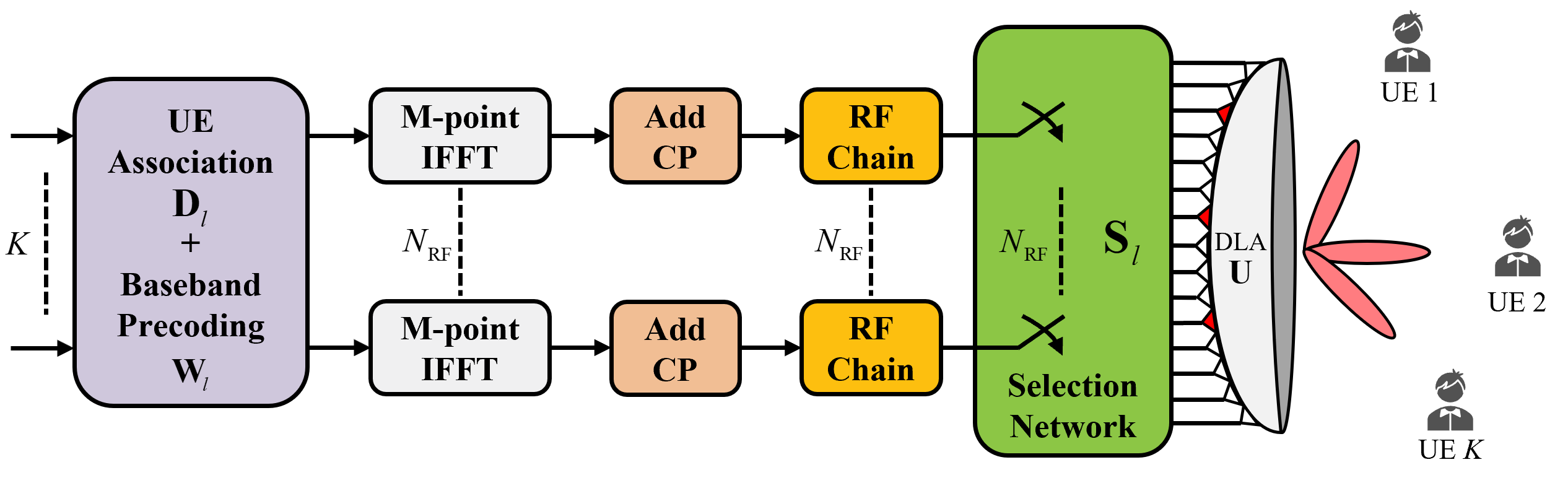}
	\caption{The typical hybrid precoding architecture of each AAU for mmWave CF-mMIMO-OFDM systems.}
	\label{figure2}
\end{figure}

The conventional hybrid precoding architecture relying on DLA of each AAU for the downlink mmWave CF-mMIMO-OFDM system is shown in Fig. \ref{figure2}. Following UE association and baseband precoding, the precoded data streams are transformed into the time domain using the inverse fast fourier transform (IFFT). Subsequently, CP is added before each OFDM symbol, and the time-domain signals are transmitted through a subset of antennas selected by a selection network. In this subsection, the subcarrier index is omitted. The precoded signal ${{\mathbf{x}}_{{\text{D}},l}} \in {\mathbb{C}^{{N_{{\text{RF}}}} \times 1}}$ transmitted by AAU $l$ is given by
\begin{equation}\label{x_D_l}
	{{\mathbf{x}}_{{\text{D}},l}} = {{\mathbf{W}}_l}{{\mathbf{D}}_l}{\mathbf{s}} = \sum\limits_{i = 1}^K {{u_{i,l}}{{\mathbf{w}}_{i,l}}{s_i}}  = \sum\limits_{i \in {\mathcal{\mathcal{D}}_l}} {{{\mathbf{w}}_{i,l}}{s_i}},
\end{equation}
where ${{\bf{W}}_l} = \left[ {{{\bf{w}}_{1,l}}, \ldots ,{{\bf{w}}_{K,l}}} \right] \in {\mathbb{C}^{{N_{{\rm{RF}}}} \times K}}$ is the digital precoding matrix at AAU $l$, where ${{\bf{w}}_{k,l}}$ is the digital precoding vector utilized for the transmission between the AAU $l$ and UE $k$. ${\mathcal{{\cal D}}_l}$ is the set of UEs served by AAU $l$. ${u_{k,l}}$ is the association indicator between AAU $l$ and UE $k$, that is, when the AAU $l$ serves UE $k$, ${u_{k,l}} = 1$ and $k \in {\mathcal{D}_l}$, otherwise, ${u_{k,l}} = 0$. The association matrix of AAU $l$ is ${{\bf{D}}_l} = {\rm{diag}}\left( {\left[ {{u_{1,l}}, \ldots ,{u_{K,l}}} \right]} \right)$. The collective received downlink signal vector ${\bf{y}} \in {\mathbb{C}^{K \times 1}}$ of all the $K$ UEs can be expressed as 
\begin{equation}\label{y_all_UE}
	{\bf{y}} = \sum\limits_{l = 1}^L {{\bf{H}}_l^{\rm{H}}{{\bf{U}}^{\rm{H}}}{{\bf{S}}_l}{{\bf{W}}_l}{{\bf{D}}_l}{\bf{s}}}  + {\bf{n}} = \sum\limits_{l = 1}^L {{\bf{\tilde H}}_l^{\rm{H}}{{\bf{S}}_l}{{\bf{W}}_l}{{\bf{D}}_l}{\bf{s}}}  + {\bf{n}}.
\end{equation}

The traditional spatial-domain channel can be converted into the beam-domain channel by employing a meticulously designed DLA \cite{gao2016fast}. This allows the antenna array to focus energy for transmitting signals on different spatial directions or receiving signals from different spatial orientations. Consequently, UEs in different directions can be distinguished. The similar effect can also be achieved using uniform linear arrays equipped with a phase shifter network. 

Specifically, such a DLA plays the role of an $N \times N$ discrete fourier transform (DFT) matrix ${\bf{U}}$, which satisfies ${{\bf{U}}^{\rm{H}}}{\bf{U}} = {\bf{I}}$. The DFT matrix ${\bf{U}}$ comprises the array steering vectors of $N$ orthogonal directions (beams) distributed across the entire angular domain as:
\begin{equation}\label{U}
	{\bf{U}} = {\left[ {{\bf{a}}\left( {{{\bar \psi }_1}} \right),{\bf{a}}\left( {{{\bar \psi }_2}} \right), \ldots ,{\bf{a}}\left( {{{\bar \psi }_N}} \right)} \right]^{\rm{H}}},
\end{equation}
where ${\bar \psi _n} = \frac{1}{N}\left( {n - \frac{{N + 1}}{2}} \right)$ are the predefined normalized spatial directions. ${\bf{a}}\left( \psi  \right) = \frac{1}{{\sqrt N }}{\left[ {{e^{ - j2\pi \psi i}}} \right]_{i \in {\cal I}}}$ represents the $N \times 1$ array response vector, where ${\cal I} = \left\{ {q - {{\left( {N - 1} \right)} \mathord{\left/{\vphantom {{\left( {N - 1} \right)} 2}} \right.\kern-\nulldelimiterspace} 2},q = 0,1, \ldots ,N - 1} \right\}$ represents the index set of array elements. The normalized spatial direction ${\psi _n}$ is related to the physical direction (angle) of propagation ${\theta _n}$, that is ${\psi _n} = \frac{d}{\lambda }\sin {\theta _n}$, where $\lambda$ is the signal wavelength and $d$ is the antenna spacing of AAU, typically chosen as $d = \frac{\lambda }{2}$. The beam-domain channel matrix from all UEs to AAU $l$ can be expressed as
\begin{equation}\label{H_beam_l}
	{{\bf{\tilde H}}_l} = \left[ {{{{\bf{\tilde h}}}_{1,l}}, \ldots ,{{{\bf{\tilde h}}}_{K,l}}} \right] = {\bf{U}}{{\bf{H}}_l} = \left[ {{\bf{U}}{{\bf{h}}_{1,l}}, \ldots ,{\bf{U}}{{\bf{h}}_{K,l}}} \right],
\end{equation}
where ${{\bf{\tilde h}}_{k,l}}$ is the beam-domain channel between AAU $l$ and UE $k$. The $N$ rows of ${{\bf{\tilde H}}_l}$ and ${{\bf{\tilde h}}_{k,l}}$ correspond to $N$ orthogonal beams and the spatial direction of each beam is ${\bar \psi _n}$. ${\bf{n}} \sim {\cal C}{\cal N}\left( {0,\sigma _{{\rm{dl}}}^2{{\bf{I}}_K}} \right)$ is a $K \times 1$ zero-mean additive white Gaussian noise (AWGN) with covariance matrix $\mathbb{E}\left\{ {{\bf{n}}{{\bf{n}}^{\rm{H}}}} \right\} = \sigma _{{\rm{dl}}}^2{{\bf{I}}_K}$, where $\sigma _{{\rm{dl}}}^2$ denotes the noise variance.

${{\mathbf{S}}_l} \in {\left\{ {0,1} \right\}^{N \times {N_{{\text{RF}}}}}}$ in (\ref{y_all_UE}) is a beam select matrix, which corresponds to the selection network in Fig. \ref{figure2}. Each column contains a single non-zero element, "1", indicating that the RF signal generated by a single RF chain is directed towards a specific beam direction. (\ref{y_all_UE}) can be further simplified as
\begin{equation}\label{y_all_UE_2}
	{\mathbf{y}} = \sum\limits_{l = 1}^L {{\mathbf{\bar H}}_l^{\text{H}}{{\mathbf{W}}_l}{{\mathbf{D}}_l}{\mathbf{s}}}  + {\mathbf{n}},
\end{equation}
where ${\mathbf{\bar H}}_l^{\text{H}} = {\mathbf{\tilde H}}_l^{\text{H}}{{\mathbf{S}}_l} \in {\mathbb{C}^{K \times {N_{{\text{RF}}}}}}$. ${{\bf{\bar H}}_l} = {{\bf{\tilde H}}_l}{\left( {i,:} \right)_{i \in {\Gamma _l}}} \in {\mathbb{C}^{\left| {{\Gamma _l}} \right| \times K}}$ is the dimension-reduced beam-domain channel matrix, which only includes the selected beams. ${\Gamma _l}$ represents the index set of selected beams by AAU $l$. We consider $\left| {{\Gamma _l}} \right| = {N_{{\text{RF}}}}$ without the loss of generality in this paper. Besides, we can also have the selected beam-domain channel vector from each UE $k$ to AAU $l$, that is ${{\bf{\bar h}}_{k,l}} = {{\bf{\bar H}}_l}\left( {:,k} \right)$.

\subsection{Channel Model}
MmWave channels are characterized by a restricted number of scattering paths. In this paper, we utilize the well-known Saleh-Valenzuela channel model for the mmWave CF-mMIMO-OFDM system. Specifically, ${{\bf{h}}_{k,l}}\left[ m \right] \in {\mathbb{C}^{N \times 1}}$ is the spatial-domain channel between AAU $l$ to UE $k$ at the $m$-th subcarrier. It can be expressed as
\begin{equation}\label{h_spatial_klm}
	{{\bf{h}}_{k,l}}\left[ m \right] = \sqrt {\frac{N}{{{\beta _{k,l}}{P_{k,l}}}}} \sum\limits_{p = 1}^{{P_{k,l}}} {\alpha _{k,l}^p{e^{ - j2\pi {\tau _{k,l,p}}{f_m}}}{\bf{a}}\left( {{\psi _{k,l,p}}} \right)},
\end{equation}
where ${P_{k,l}}$ is the number of effective channel paths corresponding to a limited number of scatters between UE $k$ and AAU $l$. $\alpha _{k,l}^p$, ${\tau _p}$ and ${\psi _{k,l,p}}$ represent the complex gain, the time delay and the angle of departure of the $p$-th path. ${f_m} = {f_c} + \frac{B}{M}\left( {m - 1 - \frac{{M - 1}}{2}} \right)$ is the frequency of the $m$-th subcarrier with ${f_c}$ and $B$ representing the carrier frequency and the bandwidth (sampling rate). ${\beta _{k,l}}$ represents the large-scale fading coefficient between AAU $l$ and UE $k$, including path loss and shadow fading in dB as
\begin{equation}\label{beta_kl}
	{\beta _{k,l}}\left[ {{\rm{dB}}} \right] = 20{\log _{10}}\left( {\frac{{4\pi {f_c}}}{c}} \right) + 10\vartheta {\log _{10}}\left( {{d_{k,l}}} \right) + {A_\varsigma },
\end{equation}
where ${d_{k,l}}$ is the distence (m) between AAU $l$ to UE $k$. $c$ is the speed (m/s) of light. $\vartheta $ is the path loss exponent. ${A_\varsigma }$ is a zero-mean Gaussian random variable with a standard deviation $\varsigma $ in dB representing the effect of shadow fading \cite{bjornson2019making}.

Accordingly, the beam-domain channel ${{\bf{\tilde h}}_{k,l}}\left[ m \right]$ between AAU $l$ and UE $k$ at the $m$-th subcarrier can be presented as
\begin{equation}\label{h_beam_klm}
	{{\bf{\tilde h}}_{k,l}}\left[ m \right] = {\bf{U}}{{\bf{h}}_{k,l}}\left[ m \right] = \sqrt {\frac{N}{{{\beta _{k,l}}{P_{k,l}}}}} \sum\limits_{p = 1}^{{P_{k,l}}} {\alpha _{k,l}^p{e^{ - j2\pi {\tau _p}{f_m}}}{\bf{\bar a}}\left( {{\psi _{k,l,p}}} \right)},
\end{equation}
where ${\bf{\bar a}}\left( {{\psi _{k,l,p}}} \right)$ is the $p$-th path component in the beamspace, which can be expressed as
\begin{equation}\label{a_beam}
	\begin{aligned}
		{\mathbf{\bar a}}\left( {{\psi _{k,l,p}}} \right) &= {\mathbf{Ua}}\left( {{\psi _p}} \right) \\ 
		&= {\left[ {{\Xi _N}\left( {{\psi _p} - {{\bar \psi }_1}} \right), \ldots ,{\Xi _N}\left( {{\psi _p} - {{\bar \psi }_N}} \right)} \right]^{\text{T}}}, \\ 
	\end{aligned} 
\end{equation}
where ${\Xi _N}\left( x \right) = \frac{{\sin N\pi x}}{{\sin \pi x}}$ is the Dirichlet sinc function \cite{gao2019wideband}. The time-domain channel in the beam domain can be expressed as
\begin{equation}\label{h_beam_kl_time}
	{{\mathbf{\tilde h}}_{k,l}}\left( t \right) = \sqrt {\frac{N}{{{\beta _{k,l}}{P_{k,l}}}}} \sum\limits_{p = 1}^{{P_{k,l}}} {\alpha _{k,l}^p{\mathbf{\bar a}}\left( {{\psi _{k,l,p}}} \right)\delta \left( {t - {\tau _p}} \right)}.
\end{equation}

\section{Downlink asynchronous transmission} \label{Downlink asynchronous reception}
In this section, we first analyze the impact of downlink asynchronous transmission in the beam domain. Subsequently, we propose a novel PBTA hybrid precoding architecture for each AAU to mitigate asynchronous interference. We derive the achievable rates under four different scenarios: synchronization, asynchronization, PBTA and small cell. Phase shift, ICI and ISI are all characterized. Additionally, two different downlink implementations of CF-mMIMO-OFDM system are considered, namely, centralized and distributed operations. Scalable MMSE and MR beam-domain precoding are employed in this work.

\subsection{Downlink Data Transmission}
\label{subsection:Downlink data transmission}
In the case of downlink asynchronous transmission, the time-domain received $a$-th OFDM symbol at UE $k$ after CP removal can be expressed as
\begin{equation}\label{z_k}
	z_k^a\left( t \right) = \sum\limits_{l = 1}^L {\left( {z_{k,l}^a\left( {t - {\delta _{k,l}}} \right){\omega _{k,l}}\left( t \right) + {j_{k,l}}\left( t \right)} \right)}  + {r_k}\left( t \right),
\end{equation}
where ${\delta _{k,l}}$ is the asynchronous timing offset\footnote{Each AAU adjusts its transmission timing according to the time reference of its nearest UE. This ensures that for every AAU, there is no delay on the received signal in the nearest UE, and also reduces the propagation delays on the subsequent UEs \cite{chowdhury2023resilient}.}. Assuming that the first arrived signal to UE $k$ is transmitted from AAU $i$ and its propagation time is ${t_{k,i}} = {{{d_{k,i}}} \mathord{\left/{\vphantom {{{d_{k,i}}} c}} \right.\kern-\nulldelimiterspace} c}$. The asynchronous timing offset in the sampling interval from AAU $l$ to UE $k$ is ${\delta _{k,l}} = \frac{{{t_{k,l}} - {t_{k,i}}}}{{{T_s}}}$, where ${T_s}$ is symbol duration. The signal of AAU $i$ reaches UE $k$ first, so ${\delta _{k,i}} = 0$. $z_{k,l}^a\left( t \right) = {\mathbf{\bar h}}_{k,l}^{\text{H}}\left( t \right){ \otimes _M}{\mathbf{x}}_{{\text{D}},l}^a\left( t \right)$ represents the received signal of UE $k$ from AAU $l$ at the $a$-th OFDM symbol in the time domain, where ${ \otimes _M}$ stands for the cyclic convolution operation with period $M$, ${\bf{x}}_{{\rm{D}},l}^a\left( t \right) \in {\mathbb{C}^{{N_{{\rm{RF}}}} \times 1}}$ is the transmitted signal of AAU $l$ at the $a$-th OFDM symbol, which can be expressed as
\begin{equation}\label{x_l_t}
	{\mathbf{x}}_{{\text{D}},l}^a\left( t \right) = \sum\limits_{m = 0}^{M - 1} {{\mathbf{x}}_{{\text{D}},l}^a\left[ m \right] \cdot \exp \left\{ {j2\pi \frac{m}{M}t} \right\}} ,t = 0,1, \ldots ,M - 1.
\end{equation}
${\mathbf{\bar h}}_{k,l}^{\text{H}}\left( t \right),t = 0, \ldots ,{T_{{\rm{D}}}} - 1$ is the downlink beam-domain channel impulse response vector between AAU $l$ and UE $k$, where ${T_{\rm{D}}}$ is the sampling length of channel delay spread. ${r_k}\left( t \right)$ is the AWGN at UE $k$ in the time domain.

\begin{figure}[!t]
	\centering
	\includegraphics[scale=0.45]{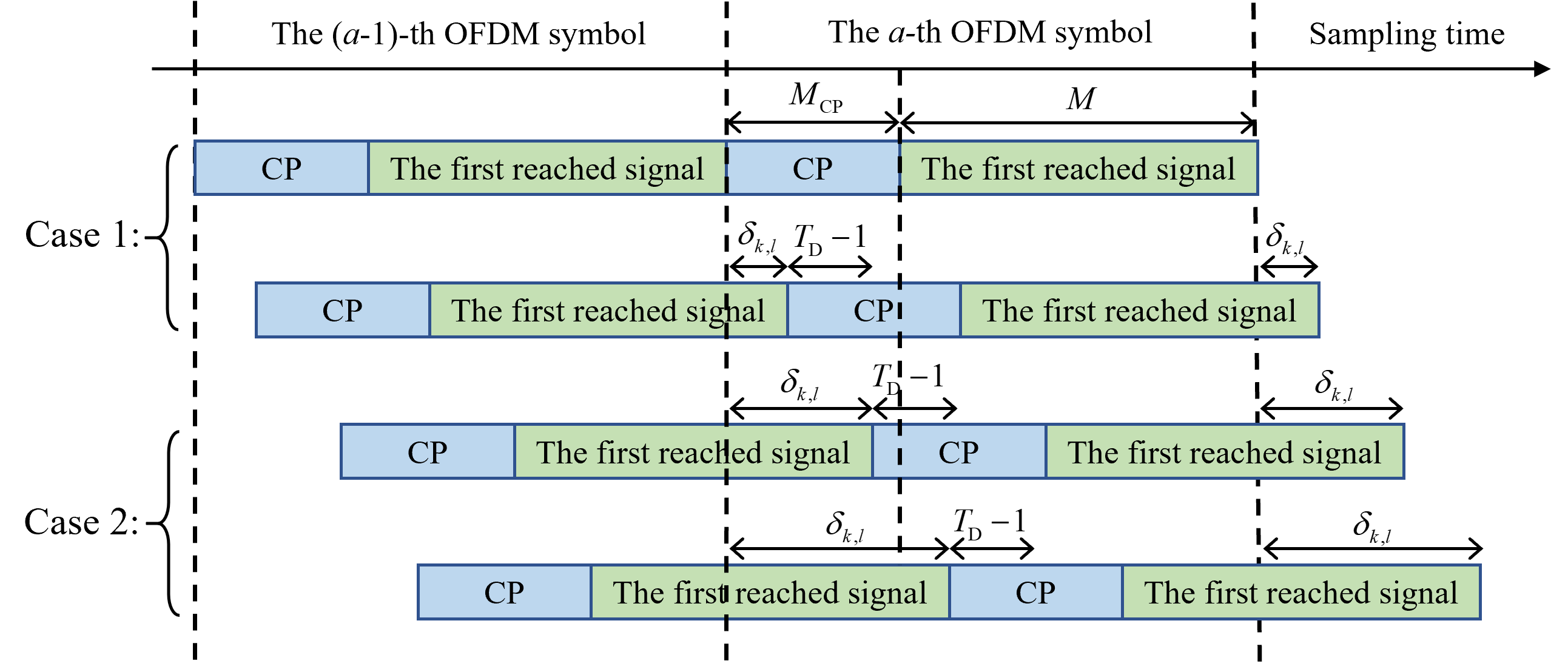}
	\caption{Two different cases for asynchronous timing offset.}
	\label{figure3}
\end{figure}

Fig. \ref{figure3} illustrates two different cases for asynchronous timing offset.
\begin{itemize}
	
	\item{Case 1: $\left( {{T_{\rm{D}}} - 1} \right) + {\delta _{k,l}} \le {M_{{\rm{CP}}}}$.} The sum of the channel delay spread and the asynchronous timing offset between AAU $l$ and UE $k$ does not exceed the length of CP. In this scenario, neither ICI nor ISI is present. Consequently, ${\omega _{k,l}}\left( t \right) = 1$ and ${j_{k,l}}\left( t \right) = 0$.
	
	\item{Case 2: $\left( {{T_{\rm{D}}} - 1} \right) + {\delta _{k,l}} > {M_{{\rm{CP}}}}$.} The sum of the channel delay spread and the asynchronous timing offset between AAU $l$ and UE $k$ exceeds the length of CP, resulting in both ICI and ISI. Similar to the system model described in \cite{yang2018pilot}, ${\omega _{k,l}}\left( t \right)$ and ${j_{k,l}}\left( t \right)$ in (\ref{z_k}) can be expressed as 
	\begin{equation}\label{omega_kl}
		{\omega _{k,l}}\left( t \right) = \begin{cases}
			0,&0 \le t \le {\delta _{k,l}} + \left( {{T_{\rm{D}}} - 1} \right) - {M_{\rm{CP}}} - 1\\
			1,&{\delta _{k,l}} + \left( {{T_{\rm{D}}} - 1} \right) - {M_{\rm{CP}}} \le t \le M - 1,
		\end{cases}
	\end{equation}
	
	\begin{small}
	\begin{equation}\label{j_kl}
		{j_{k,l}}\left( t \right) = \begin{cases}
			{\xi _{k,l}}\left( t \right),&0 \le t \le {\delta _{k,l}} + \left( {{T_{\rm{D}}} - 1} \right) - {M_{\rm{CP}}} - 1\\
			0,&{\delta _{k,l}} + \left( {{T_{\rm{D}}} - 1} \right) - {M_{\rm{CP}}} \le t \le M - 1,
		\end{cases}
	\end{equation}
	\end{small}

	where ${\rm{\xi }}_{k,l}$ is given by (\ref{xi_kl}).

\end{itemize}

\begin{figure*}[ht]
	\begin{small}
	\begin{equation}\label{xi_kl}
		\begin{array}{l}
			{{\rm{\xi }}_{k,l}}\left( t \right) = \sum\limits_{i = 0}^{t + {M_{\rm{CP}}} - {\delta _{k,l}} - 1} {{\bf{\bar h}}_{k,l}^{\rm{H}}\left( i \right){\bf{x}}_{{\rm{D}},l}^a\left( {t + M - 1 - {\delta _{k,l}} - i} \right)}
			+ \sum\limits_{i = t + {M_{\rm{CP}}} - {\delta _{k,l}}}^{{T_ {\rm{D}}} - 1} {{\bf{\bar h}}_{k,l}^{\rm{H}}\left( i \right){\bf{x}}_{{\rm{D}},l}^{a - 1}\left( {t + M - 1 + {M_{\rm{CP}}} - {\delta _{k,l}} - i} \right)}.
		\end{array}
	\end{equation}
	\end{small}
\hrulefill
\end{figure*}

\begin{figure*}[ht]
	\begin{equation}\label{y_k_1}
		y_k^a\left[ m \right] = \sum\limits_{l = 1}^L {\left( {\chi _{k,l}^m\frac{{W_{k,l}^{\left[ 0 \right]}}}{M}{\bf{\bar h}}_{k,l}^{\rm{H}}\left[ m \right]{\bf{x}}_{{\rm{D}},l}^a\left[ m \right] + \underbrace {\sum\limits_{i \ne m}^{M - 1} {\chi _{k,l}^i\frac{{W_{k,l}^{\left[ {{{\left( {m - i} \right)}_M}} \right]}}}{M}{\bf{\bar h}}_{k,l}^{\rm{H}}\left[ i \right]{\bf{x}}_{{\rm{D}},l}^a\left[ i \right]} }_{{\rm{ICI}}} + \underbrace {{\zeta _{k,l}}\left[ m \right]}_{{\rm{ISI}}}} \right)}  + {n_k}\left[ m \right].
	\end{equation}
\hrulefill
\end{figure*}
Taking $M$-point fast fourier transform (FFT) of the time-domain waveform in (\ref{z_k}), the frequency-domin received signal of UE $k$ at the $m$-th subcarrier can be expressed as (\ref{y_k_1}), where $\chi _{k,l}^m = {e^{ - j2\pi m{\delta _{k,l}}/M}}$ is the phase shift caused by asynchronous reception effect. $W_{k,l}^{\left[ m \right]}$ and ${\zeta _{k,l}}\left[ m \right]$ are respectively the FFT of ${\omega _{k,l}}\left( t \right)$ and ${j_{k,l}}\left( t \right)$. ${\bf{x}}_{{\rm{D}},l}^a\left[ m \right]$ is the frequency-domain data of AAU $l$ transmitted at the $m$-th subcarrier of the $a$-th OFDM symbol. Letting ${\varepsilon _{k,l}} = \max \left\{ {\left( {{T_{\rm{D}}} - 1} \right) + {\delta _{k,l}} - {M_{{\rm{CP}}}},0} \right\}$, $W_{k,l}^{\left[ m \right]}$ can be expressed as
\begin{equation}\label{W_kl_1}
W_{k,l}^{\left[ m \right]} = \begin{cases}
	\frac{{{e^{ - j2\pi m{\varepsilon _{k,l}}/M}} - 1}}{{1 - {e^{ - j2\pi m/M}}}},&m \ne 0\\
	M - {\varepsilon _{k,l}},&m = 0.
\end{cases}
\end{equation}

(\ref{y_k_1}) indicates that asynchronous reception will result in phase shift, ICI and ISI, due to the asynchronous timing offset ${\delta _{k,l}}$ exceeding the CP range. This issue is particularly pronounced in scenarios involving wide area coverage and ubiquitous connectivity, leading to significant performance degradation in mmWave CF-mMIMO-OFDM systems. To address this challenge, in the following subsection, we propose a novel hybrid precoding architecture called PBTA for asynchronous mmWave CF-mMIMO-OFDM systems.

\subsection{Downlink PBTA}

\begin{figure}[!t]
	\centering
	\includegraphics[scale=0.36]{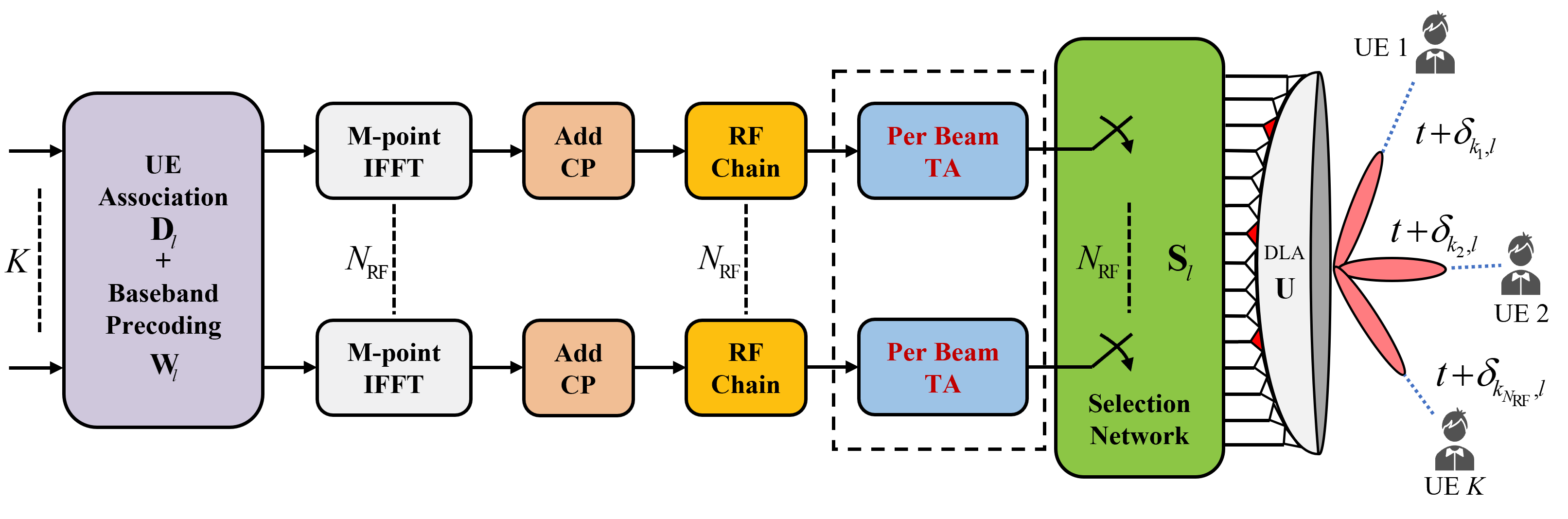}
	\caption{The proposed PBTA hybrid precoding architecture of each AAU for asynchronous mmWave CF-mMIMO-OFDM systems.}
	\label{figure4}
\end{figure}

As discussed in Subsection \ref{subsection:Downlink data transmission}, due to the asynchronous reception, the phase shift, ICI and ISI caused by asynchronous timing offset will result in severe performance degradation in mmWave CF-mMIMO-OFDM systems with conventional hybrid precoding architecture. In this section, we propose a novel hybrid precoding architecture called PBTA to solve this problem. 

As shown in Fig. \ref{figure4}, compared with the conventional hybrid precoding architecture, per-beam timing advance units are introduced as an additional precoding layer before the selection network. Specifically, each RF chain is connected to a PBTA unit, which adjusts the transmitting timing of each beam to compensate for asynchronous delays. This adjustment can change the asynchronous timing offset of the desired signal from Case 2 to Case 1. 

By employing a unified unitary matrix transformation to convert the spatial domain signal into the beam domain, the DLA can efficiently focus energy on transmitting the signal in various spatial directions. Consequently, each RF chain is allocated to a specific beam direction, with each direction corresponding to a distinct UE. This allows the transmission time of the beam directed to each UE to be adjusted, ensuring that beams assigned to the same UE from different AAUs arrive simultaneously, thereby minimizing the asynchronous timing offsets of the desired signals.

It is assumed that the AAU can obtain the information of asynchronous timing offsets from each UE to all AAUs in advance. To be more specific, during the initial stage, the AAU estimates the latency of each UE according to the uplink physical random access channel (PRACH) and sends the information to the CPU. After that, the CPU calculates the exact latency of each UE based on the transmit advance of each UE. Then the CPU calculates all asynchronous timing offsets and feeds them back to each AAU. 

To meet the synchronization criteria, the AAU calculates the TA for selected beams, which in turn performs PBTA to compensate for the asynchronous timing offset at the transmitting side. Let $i$-th RF chain be directed towards UE ${k_i}$. The transmitted signal of AAU $l$ at the $a$-th OFDM symbol after performing PBTA can be expressed as
\begin{equation}\label{x_l_t_PBTA}
	{\bf{x}}_{{\rm{D}},l}^{a\text{-}\rm{PBTA}}\left( t \right) = \left[ \begin{array}{c}
		x_{{\rm{D}},l,1}^a\left( {t + {\delta _{{k_1},l}}} \right)\\
		x_{{\rm{D}},l,2}^a\left( {t + {\delta _{{k_2},l}}} \right)\\
		\vdots \\
		x_{{\rm{D}},l,{N_{{\rm{RF}}}}}^a\left( {t +  + {\delta _{{k_{{N_{\rm{RF}}}}},l}}} \right)
	\end{array} \right].
\end{equation}
Substituting (\ref{x_l_t_PBTA}) into (\ref{z_k}) and taking $M$-point FFT, the frequency-domain received signal of UE $k$ after performing PBTA at the $m$-th subcarrier can be expressed as (\ref{y_k_2}),
\begin{figure*}[ht]
	\begin{equation}\label{y_k_2}
		y_k^a\left[ m \right] = \sum\limits_{l = 1}^L {\left( {\sum\limits_{n = 1}^{{N_{{\rm{RF}}}}} {\chi _{k,l,n}^m\frac{{W_{k,l,n}^{\left[ 0 \right]}}}{M}\bar h_{k,l,n}^{\rm{H}}\left[ m \right]x_{{\rm{D}},l,n}^a\left[ m \right]}  + \underbrace {\sum\limits_{i \ne m}^{M - 1} {\sum\limits_{n = 1}^{{N_{{\rm{RF}}}}} {\chi _{k,l,n}^i\frac{{W_{k,l,n}^{\left[ {{{\left( {m - i} \right)}_M}} \right]}}}{M}\bar h_{k,l,n}^{\rm{H}}\left[ i \right]x_{{\rm{D}},l,n}^a\left[ i \right]} } }_{{\rm{ICI}}} + \underbrace {{\zeta _{k,l}}\left[ m \right]}_{{\rm{ISI}}}} \right)}  + {n_k}\left[ m \right].
	\end{equation}
	\hrulefill
\end{figure*}
where ${x_{{\rm{D}},l,n}^a\left[ m \right]}$ is the $n$-th element of ${{\bf{x}}_{{\rm{D}},l}^a\left[ m \right]}$, which represent the $n$-th beam of AAU $l$. ${\bar h_{k,l,n}\left[ m \right]}$ is the $n$-th element of ${{{{\bf{\bar h}}}_{k,l}}\left[ m \right]}$. $\chi _{k,l,n}^m = {e^{ - j2\pi m\left( {{\delta _{k,l}} - {\delta _{{k_n},l}}} \right)/M}}$ is the phase shift of the $n$-th beam. If UE $k$ is served by $n$-th RF chain of AAU $l$, ${k_n} = k$ and ${\delta _{k,l}} = {\delta _{{k_n},l}}$. In this case, the desired signal transmitted by AAU $l$ to UE $k$ is not affected by the phase shift caused by asynchronous reception, and $\chi _{k,l,n}^m = 1$. Since the desired signal of each UE arrives synchronously, the corresponding ICI and ISI caused by asynchronous reception will also be mitigated. $W_{k,l,n}^{\left[ m \right]}$ can be expressed as
\begin{equation}\label{W_kln}
	W_{k,l,n}^{\left[ m \right]} = \begin{cases}
		\frac{{{e^{ - j2\pi m{\varepsilon _{k,l,n}}/M}} - 1}}{{1 - {e^{ - j2\pi m/M}}}},&m \ne 0\\
		M - {\varepsilon _{k,l,n}},&m = 0,
	\end{cases}
\end{equation}
where ${\varepsilon _{k,l,n}} = \max \left\{ {\left( {{T_\text{D}} - 1} \right) + {\delta _{k,l,n}} - {M_{\text{CP}}},0} \right\}$, and ${\delta _{k,l,n}} = {\delta _{k,l}} - {\delta _{{k_n},l}}$ is the asynchronous timing offset of the $n$-th beam from AAU $l$ to UE $k$. ${j_{k,l}}\left( t \right) = \sum\limits_{n = 1}^{{N_{{\rm{RF}}}}} {{j_{k,l,n}}\left( t \right)}$ represents the data that are erroneously processed, where ${j_{k,l,n}}\left( t \right)$ is the data at the $n$-th beam, which can be expressed as

\begin{small}
\begin{equation}\label{j_kln}
	{j_{k,l,n}}\left( t \right) = \begin{cases}
		{\xi _{k,l,n}}\left( t \right),&0 \le t \le {\delta _{k,l,n}} + \left( {{T_{\rm{D}}} - 1} \right) - {M_{{\rm{CP}}}} - 1\\
		0,&{\delta _{k,l,n}} + \left( {{T_{\rm{D}}} - 1} \right) - {M_{{\rm{CP}}}} \le t \le M - 1,
	\end{cases}
\end{equation}
\end{small}%
where ${\xi _{k,l,n}}\left( t \right)$ can be expressed as (\ref{xi_kln}).
\begin{figure*}[ht]
	\begin{small}
	\begin{equation}\label{xi_kln}
		{{\rm{\xi }}_{k,l,n}}\left( t \right) = \sum\limits_{i = 0}^{t + {M_{{\rm{CP}}}} - {\delta _{k,l,n}} - 1} {\bar h_{k,l,n}^{\rm{H}}\left( i \right)x_{{\rm{D}},l,n}^a\left( {t + M - 1 - {\delta _{k,l,n}} - i} \right)}  + \sum\limits_{i = t + {M_{{\rm{CP}}}} - {\delta _{k,l,n}}}^{{T_{\rm{D}}} - 1} {\bar h_{k,l,n}^{\rm{H}}\left( i \right)x_{{\rm{D}},l,n}^{a - 1}\left( {t + M - 1 + {M_{{\rm{CP}}}} - {\delta _{k,l,n}} - i} \right)}.
	\end{equation}
	\end{small}
	\hrulefill
\end{figure*}
Let $\kappa _{k,l,n,m}^i = \chi _{k,l,n}^i\frac{{W_{k,l,n}^{\left[ {{{\left( {m - i} \right)}_M}} \right]}}}{M}$, where $\kappa _{k,l,n,m}^i$ encapsulates the phase shift and amplitude attenuation of the $n$-th beam from AAU $l$ to UE $k$ caused by asynchronous reception at the $m$-th subcarrier. Consequently, (\ref{y_k_2}) can be further simplified to (\ref{y_k_3}). We define the phase shift diagonal matrix of AAU $l$ to UE $k$ at the $m$-th subcarrier as $\Theta _{k,l,m}^i \buildrel \Delta \over = {\rm{diag}}\left( {\kappa _{k,l,1,m}^i, \ldots ,\kappa _{k,l,{N_{{\rm{RF}}}},m}^i} \right) \in {\mathbb{C}^{{N_{{\rm{RF}}}} \times {N_{{\rm{RF}}}}}}$, and substitute it into (\ref{y_k_3}). This yields a more concise result, as shown in (\ref{y_k_4}).
\begin{figure*}[ht]
	\begin{equation}\label{y_k_3}
		y_k^a\left[ m \right] = \sum\limits_{l = 1}^L {\left( {\sum\limits_{n = 1}^{{N_{{\rm{RF}}}}} {\kappa _{k,l,n,m}^m\bar h_{k,l,n}^{\rm{H}}\left[ m \right]x_{{\rm{D}},l,n}^a\left[ m \right]}  + \underbrace {\sum\limits_{i \ne m}^{M - 1} {\sum\limits_{n = 1}^{{N_{{\rm{RF}}}}} {\kappa _{k,l,n,m}^i\bar h_{k,l,n}^{\rm{H}}\left[ i \right]x_{{\rm{D}},l,n}^a\left[ i \right]} } }_{{\rm{ICI}}} + \underbrace {{\zeta _{k,l}}\left[ m \right]}_{{\rm{ISI}}}} \right)}  + {n_k}\left[ m \right].
	\end{equation}
	\hrulefill
\end{figure*}

\begin{figure*}[ht]
	\begin{equation}\label{y_k_4}
	y_k^a\left[ m \right] = \sum\limits_{l = 1}^L {\left( {\underbrace {{{\left( {\Theta _{k,l,m}^m{{{\bf{\bar h}}}_{k,l}}\left[ m \right]} \right)}^{\rm{H}}}{\bf{x}}_{{\rm{D}},l}^a\left[ m \right]}_{{\rm{Received \; signal}}} + \underbrace {\sum\limits_{i \ne m}^{M - 1} {{{\left( {\Theta _{k,l,m}^i{{{\bf{\bar h}}}_{k,l}}\left[ i \right]} \right)}^{\rm{H}}}{\bf{x}}_{{\rm{D}},l}^a\left[ i \right]} }_{{\rm{ICI}}} + \underbrace {{\zeta _{k,l}}\left[ m \right]}_{{\rm{ISI}}}} \right)}  + \underbrace {{n_k}\left[ m \right]}_{{\rm{Noise}}}.
	\end{equation}
	\hrulefill
\end{figure*}
For the received signal in (\ref{y_k_4}), we can deduce as follows:
\begin{equation}\label{y_k_reveive}
	\begin{aligned}
		y_{k,{\rm{r}}}^a\left[ m \right] 
		&= \sum\limits_{l = 1}^L {{{\left( {\Theta _{k,l,m}^m{{{\bf{\bar h}}}_{k,l}}\left[ m \right]} \right)}^{\rm{H}}}{\bf{x}}_{{\rm{D}},l}^a\left[ m \right]}\\
		&= \sum\limits_{l = 1}^L {{{\left( {\Theta _{k,l,m}^m{{{\bf{\bar h}}}_{k,l}}\left[ m \right]} \right)}^{\rm{H}}}\left( {\sum\limits_{i = 1}^K {{u_{i,l}}{{\bf{w}}_{i,l}}\left[ m \right]{s_i}\left[ m \right]} } \right)}\\
		&= \sum\limits_{i = 1}^K {{{\left[ \begin{array}{c}
						\Theta _{k,1,m}^m{{{\bf{\bar h}}}_{k,1}}\left[ m \right]\\
						\vdots \\
						\Theta _{k,L,m}^m{{{\bf{\bar h}}}_{k,L}}\left[ m \right]
					\end{array} \right]}^{\rm{H}}}\left[ \begin{array}{l}
				{u_{i,1}}{{\bf{w}}_{i,1}}\left[ m \right]\\
				\vdots \\
				{u_{i,L}}{{\bf{w}}_{i,L}}\left[ m \right]
			\end{array} \right]} {s_i}\left[ m \right]\\
		&= \sum\limits_{i = 1}^K {{{\left( {\Theta _{k,m}^m{{{\bf{\bar h}}}_k}\left[ m \right]} \right)}^{\rm{H}}}{{\bf{D}}_i}{{\bf{w}}_i}\left[ m \right]{s_i}\left[ m \right]}\\
		&= \underbrace {{{\left( {\Theta _{k,m}^m{{{\bf{\bar h}}}_k}\left[ m \right]} \right)}^{\rm{H}}}{{\bf{D}}_k}{{\bf{w}}_k}\left[ m \right]{s_k}\left[ m \right]}_{{\rm{Desired \; signal}}} + \\ &\underbrace {\sum\limits_{i = 1,i \ne k}^K {{{\left( {\Theta _{k,m}^m{{{\bf{\bar h}}}_k}\left[ m \right]} \right)}^{\rm{H}}}{{\bf{D}}_i}{{\bf{w}}_i}\left[ m \right]{s_i}\left[ m \right]} }_{{\rm{Inter \text{-} user \; interference}}},
	\end{aligned}
\end{equation}
where $\Theta _{k,m}^m = {\rm{blkdiag}}\left( {\Theta _{k,1,m}^m, \ldots ,\Theta _{k,L,m}^m} \right) \in {\mathbb{C}^{L{N_{{\rm{RF}}}} \times L{N_{{\rm{RF}}}}}}$ denotes the phase shift block diagonal matrix of UE $k$ to all AAUs at the $m$-th subcarrier. ${{\bf{\bar h}}_k}\left[ m \right] = {\left[ {{\bf{\bar h}}_{k,1}^{\rm{T}}\left[ m \right], \ldots ,{\bf{\bar h}}_{k,L}^{\rm{T}}\left[ m \right]} \right]^{\rm{T}}} \in {\mathbb{C}^{L{N_{{\rm{RF}}}} \times 1}}$ represents the collective beam-domain channel vector of UE $k$ at the $m$-th subcarrier. ${{\bf{D}}_i} = {\rm{diag}}\left( {{u_{i,1}}, \ldots ,{u_{i,L}}} \right) \otimes {{\bf{I}}_{{N_{{\rm{RF}}}}}} \in {\left\{ {0,1} \right\}^{L{N_{{\rm{RF}}}} \times L{N_{{\rm{RF}}}}}}$ denotes the block assoiation matrix of UE $i$. ${{\bf{w}}_i}\left[ m \right] = {\left[ {{\bf{w}}_{i,1}^{\rm{T}}\left[ m \right], \ldots ,{\bf{w}}_{i,L}^{\rm{T}}\left[ m \right]} \right]^{\rm{T}}} \in {\mathbb{C}^{L{N_{{\rm{RF}}}} \times 1}}$ is the collective precoding vector of UE $i$ at the $m$-th subcarrier.

For ICI in (\ref{y_k_4}), we can deduce as follows:
\begin{equation}\label{y_k_ICI}
	\begin{aligned}
		y_{k,{\rm{ICI}}}^a\left[ m \right] 
		&= \sum\limits_{l = 1}^L {\sum\limits_{i \ne m}^{M - 1} {{{\left( {\Theta _{k,l,m}^i{{{\bf{\bar h}}}_{k,l}}\left[ i \right]} \right)}^{\rm{H}}}{\bf{x}}_{{\rm{D}},l}^a\left[ i \right]} } \\
		&= \sum\limits_{i \ne m}^{M - 1} {\sum\limits_{l = 1}^L {{{\left( {\Theta _{k,l,m}^i{{{\bf{\bar h}}}_{k,l}}\left[ i \right]} \right)}^{\rm{H}}}\left( {\sum\limits_{i = 1}^K {{u_{i,l}}{{\bf{w}}_{i,l}}\left[ i \right]{s_i}\left[ i \right]} } \right)} } \\
		&= \sum\limits_{i \ne m}^{M - 1} {\sum\limits_{j = 1}^K {{{\left( {\Theta _{k,m}^i{{{\bf{\bar h}}}_k}\left[ i \right]} \right)}^{\rm{H}}}{{\bf{D}}_j}{{\bf{w}}_j}\left[ i \right]{s_j}\left[ i \right]} }.
	\end{aligned}
\end{equation}
Based on (\ref{y_k_4}), (\ref{y_k_reveive}) and (\ref{y_k_ICI}), the downlink effective signal-to-interference-and-noise ratio (SINR) of UE $k$ at the $m$-th subcarrier can be expressed as (\ref{SINR_k_asyn}).
\begin{figure*}[ht]
	\begin{equation}\label{SINR_k_asyn}
		\gamma _k^{{\rm{Asyn\text{-}dl}}}\left[ m \right] = \frac{{{{\left| {{{\left( {\Theta _{k,m}^m{{{\bf{\bar h}}}_k}\left[ m \right]} \right)}^{\rm{H}}}{{\bf{D}}_k}{{\bf{w}}_k}\left[ m \right]} \right|}^2}}}{{\sum\limits_{i = 1,i \ne k}^K {{{\left| {{{\left( {\Theta _{k,m}^m{{{\bf{\bar h}}}_k}\left[ m \right]} \right)}^{\rm{H}}}{{\bf{D}}_i}{{\bf{w}}_i}\left[ m \right]} \right|}^2}}  + \sum\limits_{i \ne m}^{M - 1} {\sum\limits_{j = 1}^K {{{\left| {{{\left( {\Theta _{k,m}^i{{{\bf{\bar h}}}_k}\left[ i \right]} \right)}^{\rm{H}}}{{\bf{D}}_j}{{\bf{w}}_j}\left[ i \right]} \right|}^2}} }  + \sum\limits_{l = 1}^L {{{\left| {{\zeta _{k,l}}\left[ m \right]} \right|}^2}}  + \sigma _{{\rm{dl}}}^2}}.
	\end{equation}
	\hrulefill
\end{figure*}

For comparison and analysis, in the hypothetical synchronization scenario, the downlink SINR of UE $k$ at the $m$-th subcarrier can be expressed as
\begin{equation}\label{SINR_k_syn}
	\gamma _k^{{\rm{Syn\text{-}dl}}}\left[ m \right] = \frac{{{{\left| {{\bf{\bar h}}_k^{\rm{H}}\left[ m \right]{{\bf{D}}_k}{{\bf{w}}_k}\left[ m \right]} \right|}^2}}}{{\sum\limits_{i = 1,i \ne k}^K {{{\left| {{\bf{\bar h}}_k^{\rm{H}}\left[ m \right]{{\bf{D}}_i}{{\bf{w}}_i}\left[ m \right]} \right|}^2}}  + \sigma _{{\rm{dl}}}^2}}.
\end{equation}
The achievable downlink SE of UE $k$ at the $m$-th subcarrier is given by
\begin{equation}\label{SE_k}
	{R_k^{\text{dl}}}\left[ m \right] = \left( {\frac{M}{{M + {M_{{\rm{CP}}}}}}} \right)\mathbb{E}\left\{ {{{\log }_2}\left( {1 + {\gamma _k}\left[ m \right]} \right)} \right\}.
\end{equation}

The small-cell network is the special implementation. Each UE is exclusively served by one AAU, so there is no interference of asynchronization in practice. Therefore, we use it as an actual benchmark scenario. In this case, the precoding can be performed locally at the AAU by using its own local channel without exchanging anything with the CPU. Assuming that AAU $l$ serves UE $k$, the downlink SINR of UE $k$ at the $m$-th subcarrier can be expressed as
\begin{equation}\label{SINR_kl_cellular}
	\gamma _{k,l}^{{\rm{s\text{-}dl}}}\left[ m \right] = \frac{{{{\left| {{\bf{\bar h}}_{k,l}^{\rm{H}}\left[ m \right]{{\bf{w}}_{k,l}}\left[ m \right]} \right|}^2}}}{{\sum\limits_{i = 1,i \ne k}^K {{{\left| {{\bf{\bar h}}_{k,l}^{\rm{H}}\left[ m \right]{{\bf{w}}_{i,l}}\left[ m \right]} \right|}^2}}  + \sigma _{{\rm{dl}}}^2}}.
\end{equation}
In the small-cell scenario, the achievable downlink SE of UE $k$ at the $m$-th subcarrier is given by
\begin{equation}\label{SE_kl_cellular}
	R_k^{{\rm{s\text{-}dl}}}\left[ m \right] = \left( {\frac{M}{{M + {M_{{\rm{CP}}}}}}} \right)\mathop {\max }\limits_{l \in \left\{ {1, \ldots ,L} \right\}} \mathbb{E}\left\{ {{{\log }_2}\left( {1 + \gamma _{k,l}^{{\rm{s\text{-}dl}}}\left[ m \right]} \right)} \right\}.
\end{equation}

\subsection{Digital Precoders for Asynchronous Transmission}
In the numerical evaluation, we will employ the scalable MR and MMSE digital precoding in both centralized and distributed implementations \cite{bjornson2020scalable}. The downlink centralized precoding is jointly computed at CPU and defined as ${{\bf{w}}_k}\left[ m \right] = \sqrt {{\rho _k}} {{{{{\bf{\bar w}}}_k}\left[ m \right]} \mathord{\left/{\vphantom {{{{{\bf{\bar w}}}_k}\left[ m \right]} {\sqrt {\mathbb{E}\left\{ {{{\left\| {{{{\bf{\bar w}}}_k}\left[ m \right]} \right\|}^2}} \right\}} }}} \right.\kern-\nulldelimiterspace} {\sqrt {\mathbb{E}\left\{ {{{\left\| {{{{\bf{\bar w}}}_k}\left[ m \right]} \right\|}^2}} \right\}} }}$, where ${\rho _k}$ is the downlink transmission power assigned by each AAU to UE $k$, and ${\rho _{\max }}$ is the total downlink transmission power of each AAU. ${\sqrt {\mathbb{E}\left\{ {{{\left\| {{{{\bf{\bar w}}}_k}\left[ m \right]} \right\|}^2}} \right\}} }$ is the power normalization coefficient, ensuring that $\mathbb{E}\left\{ {{{\left\| {{{\bf{w}}_k}} \right\|}^2}} \right\} = {\rho _k}$. We consider a simple equal power allocation with ${\rho _k} = {{{\rho _{\max }}} \mathord{\left/{\vphantom {{{\rho _{\max }}} {{N_{{\rm{RF}}}}}}} \right.\kern-\nulldelimiterspace} {{N_{{\rm{RF}}}}}}$. 

We assume that the delay phase can be perfectly known by positioning or other technologies \cite{zheng2023asynchronous}, and use the delay phase used (DU) precoding. In practice, the estimated channel is also affected by asynchronous reception, so the asynchronous phase shift should be incorporated into the channel used for calculating precoding. This approach mitigates the impact of phase shifts, ensuring that the asynchronous phase offsets at different subcarriers have little impact on our subsequent simulations. We assume perfect channel state information (CSI) at the transmitter, allowing us to focus on investigating the impacts of asynchronous reception without the influence of channel estimation errors. ${{\bf{\bar w}}_k}\left[ m \right]$ for MR can be expressed as ${{\bf{\bar w}}_k}^{{\rm{MR}}}\left[ m \right]={{\bf{D}}_k}\Theta _{k,m}^m{{{\bf{\bar h}}}_k}\left[ m \right]$ and for scalable partial MMSE (P-MMSE) can be expressed as (\ref{w_mmse_c}).
\begin{figure*}[ht]
	\begin{equation}\label{w_mmse_c}
		{{\bf{\bar w}}_k}^{{\rm{P\text{-}MMSE}}}\left[ m \right] = {p_k}{\left( {\sum\limits_{i = 1}^K {{p_i}{{\bf{D}}_k}\Theta _{i,m}^m{{{\bf{\bar h}}}_i}\left[ m \right]{{\left( {\Theta _{i,m}^m{{{\bf{\bar h}}}_i}\left[ m \right]} \right)}^{\rm{H}}}{{\bf{D}}_k}}  + {\sigma ^2}{{\bf{D}}_k}} \right)^{ - 1}}{{\bf{D}}_k}\Theta _{k,m}^m{{{\bf{\bar h}}}_k}\left[ m \right].
	\end{equation}
	\hrulefill
\end{figure*}

For distributed implementation, precoding vectors are computed at each AAU with the CSIs of the UEs associated with the AAU. The downlink distributed precoding is defined as ${{\bf{w}}_{k,l}}\left[ m \right] = \sqrt {{\rho _k}} {{{{{\bf{\bar w}}}_{k,l}}\left[ m \right]} \mathord{\left/{\vphantom {{{{{\bf{\bar w}}}_{k,l}}\left[ m \right]} {\sqrt {\mathbb{E}\left\{ {{{\left\| {{{{\bf{\bar w}}}_{k,l}}\left[ m \right]} \right\|}^2}} \right\}} }}} \right.\kern-\nulldelimiterspace} {\sqrt {\mathbb{E}\left\{ {{{\left\| {{{{\bf{\bar w}}}_{k,l}}\left[ m \right]} \right\|}^2}} \right\}} }}$, where ${{\bf{\bar w}}_{k,l}}\left[ m \right]$ for local MR (L-MR) can be expressed as ${{\bf{\bar w}}_{k,l}}^{{\rm{L\text{-}MR}}}\left[ m \right]=\Theta _{k,l,m}^m{{{\bf{\bar h}}}_{k,l}}\left[ m \right]$ and for scalable local partial MMSE (LP-MMSE) can be expressed as (\ref{w_mmse_d}).
\begin{figure*}[ht]
	\begin{equation}\label{w_mmse_d}
		{\bf{\bar w}}_{k,l}^{{\rm{LP\text{-}MMSE}}}\left[ m \right] = {p_k}{\left( {\sum\limits_{i \in {{\cal D}}_l} {{p_i}\left( {\Theta _{i,l,m}^m{{{\bf{\bar h}}}_{i,l}}\left[ m \right]{{\left( {\Theta _{i,l,m}^m{{{\bf{\bar h}}}_{i,l}}\left[ m \right]} \right)}^{\rm{H}}}} \right)}  + {\sigma ^2}{{\bf{I}}_{{N_{{\rm{RF}}}}}}} \right)^{ - 1}}\Theta _{k,l,m}^m{{{\bf{\bar h}}}_{k,l}}\left[ m \right].
	\end{equation}
	\hrulefill
\end{figure*}

\section{Uplink Asynchronous Reception} \label{Uplink Asynchronous reception}
In this section, we analyze the impact of uplink asynchronous reception. Subsequently, we apply the proposed PBTA hybrid precoding architecture to the uplink. We derive the achievable rates of four scenarios and consider two distinct implementations, including centralized and distributed operations.
\subsection{Uplink Data Reception}
In the case of uplink asynchronous reception, the time-domain received $a$-th OFDM symbol at AAU $l$ after CP removal is ${\bf{z}}_l^a\left( t \right) \in {\mathbb{C}^{{N_{{\rm{RF}}}} \times 1}}$, which can be expressed as
\begin{equation}\label{z_l}
	{\bf{z}}_l^a\left( t \right) = \sum\limits_{k = 1}^K {\sqrt {{p_k}} \left( {{\bf{z}}_{k,l}^a\left( {t - {\delta _{k,l}}} \right){\omega _{k,l}}\left( t \right) + {{\bf{j}}_{k,l}}\left( t \right)} \right)}  + {{\bf{r}}_l}\left( t \right),
\end{equation}
where ${\bf{z}}_{k,l}^a\left( t \right) = {{\bf{\bar h}}_{k,l}}\left( t \right){ \otimes _M}s_k^a\left( t \right)$, $s_k^a\left( t \right)$ is the uplink complex data symbol transmitted by UE $k$ at the $a$-th OFDM symbol. ${p_k}$ is the uplink transmit power of UE $k$. ${\omega _{k,l}}\left( t \right)$ is defined as in (\ref{omega_kl}). ${{\bf{r}}_l}\left( t \right) \in {\mathbb{C}^{{N_{{\rm{RF}}}} \times 1}}$ is the AWGN with variance $\sigma _{{\text{ul}}}^2$ at AAU $l$ in the time domain. ${{\bf{j}}_{k,l}}\left( t \right) \in {\mathbb{C}^{{N_{{\rm{RF}}}} \times 1}}$ is the data that are erroneously processed from UE $k$ to AAU $l$, which can be expressed as 
\begin{equation}\label{j_kl_2}
{{\bf{j}}_{k,l}}\left( t \right) =  \begin{cases}
	{{\boldsymbol{\xi }}_{k,l}}\left( t \right),&0 \le t \le {\delta _{k,l}} + \left( {{T_{\rm{D}}} - 1} \right) - {M_{{\rm{CP}}}} - 1\\
	0,&{\delta _{k,l}} + \left( {{T_{\rm{D}}} - 1} \right) - {M_{{\rm{CP}}}} \le t \le M - 1,
\end{cases}
\end{equation}
where
\begin{equation}\label{xi_kl_2}
	\begin{array}{l}
		{{\boldsymbol{\xi }}_{k,l}}\left( t \right) = \sum\limits_{i = 0}^{t + {M_{{\rm{CP}}}} - {\delta _{k,l}} - 1} {{{{\bf{\bar h}}}_{k,l}}\left( i \right)s_k^a\left( {t + M - 1 - {\delta _{k,l}} - i} \right)} \\
		+ \sum\limits_{i = t + {M_{{\rm{CP}}}} - {\delta _{k,l}}}^{{T_{\rm{D}}} - 1} {{{{\bf{\bar h}}}_{k,l}}\left( i \right)s_k^{a - 1}\left( {t + M - 1 + {M_{{\rm{CP}}}} - {\delta _{k,l}} - i} \right)}.
	\end{array}
\end{equation}

\subsection{Uplink PBTA}
For uplink reception, AAUs utilizing DLAs can effectively gather energy to receive signals from UEs in different directions. Unlike the downlink scenario, we propose implementing PBTA at the receiving side. We assume that each AAU can obtain the CSI and the asynchronous timing offset of each UE it serves. Subsequently, each AAU adjusts the timing advance of each beam according to the asynchronous timing offset of each UE. This ensures that each receiving beam is synchronized with its corresponding UE, allowing the desired signals from different UEs to reach each AAU simultaneously. 

After performing PBTA, the adjusted signal of ${\bf{z}}_{k,l}^{a}\left( t \right)$ in (\ref{z_l}) is given by

\begin{small}
\begin{equation}\label{z_kl}
	{\bf{z}}_{k,l}^{a\text{-PBTA}}\left( t \right) = \left[ \begin{array}{c}
		z_{k,l,1}^a\left( {t + {\delta _{{k_1},l}}} \right)\\
		z_{k,l,2}^a\left( {t + {\delta _{{k_2},l}}} \right)\\
		\vdots \\
		z_{k,l,{N_{{\rm{RF}}}}}^a\left( {t + {\delta _{{k_{{N_{RF}}}},l}}} \right)
	\end{array} \right],
\end{equation}
\end{small}%
where $z_{k,l,n}^a\left( {t + {\delta _{{k_n},l}}} \right)$ is the adjusted reveived signal at the $n$-th beam. (\ref{z_l}) can be represented as (\ref{z_l_2}), where ${\xi _{k,l,n}}\left( t \right)$ in ${j_{k,l,n}}\left( t \right)$ can be expressed as

\begin{figure*}[ht]
	\begin{small}
	\begin{equation}\label{z_l_2}
	{\bf{z}}_l^a\left( t \right) = \sum\limits_{k = 1}^K {\sqrt {{p_k}} \left( {\left[ \begin{array}{c}
				z_{k,l,1}^a\left( {t + {\delta _{{k_1},l}} - {\delta _{k,l}}} \right)\\
				z_{k,l,2}^a\left( {t + {\delta _{{k_2},l}} - {\delta _{k,l}}} \right)\\
				\vdots \\
				z_{k,l,{N_{{\rm{RF}}}}}^a\left( {t + {\delta _{{k_{{N_{RF}}}},l}} - {\delta _{k,l}}} \right)
			\end{array} \right] \odot \left[ \begin{array}{c}
				{\omega _{k,l,1}}\left( t \right)\\
				{\omega _{k,l,2}}\left( t \right)\\
				\vdots \\
				{\omega _{k,l,{N_{{\rm{RF}}}}}}\left( t \right)
			\end{array} \right] + \left[ \begin{array}{c}
				{j_{k,l,1}}\left( t \right)\\
				{j_{k,l,2}}\left( t \right)\\
				\vdots \\
				{j_{k,l,{N_{{\rm{RF}}}}}}\left( t \right)
			\end{array} \right]} \right)}  + {{\bf{r}}_l}\left( t \right),
	\end{equation}
	\hrulefill
	\end{small}%
\end{figure*}

\begin{small}
\begin{equation}\label{xi_kln_2}
	\begin{array}{l}
		{\xi _{k,l,n}}\left( t \right) = \sum\limits_{i = 0}^{t + {M_{{\rm{CP}}}} - {\delta _{k,l,n}} - 1} {{{\bar h}_{k,l,n}}\left( i \right)s_k^a\left( {t + M - 1 - {\delta _{k,l,n}} - i} \right)} \\
		+ \sum\limits_{i = t + {M_{{\rm{CP}}}} - {\delta _{k,l,n}}}^{{T_{\rm{D}}} - 1} {{{\bar h}_{k,l,n}}\left( i \right)s_k^{a - 1}\left( {t + M - 1 + {M_{{\rm{CP}}}} - {\delta _{k,l,n}} - i} \right)}.
	\end{array}
\end{equation}
\end{small}%

Similarly, as in the downlink scenario, by performing $M$-point FFT on (\ref{z_l_2}), the received signal at the AAU $l$ and the collective received signals at the CPU at the $m$-th subcarrier can be derived as (\ref{y_l}) and (\ref{y}), where ${\bf{y}}_l^a\left[ m \right]$ denotes the received signals of AAU $l$ and ${{\bf{y}}^a}\left[ m \right] = {\left[ {{{\left( {{\bf{y}}_1^a\left[ m \right]} \right)}^{\rm{T}}},{{\left( {{\bf{y}}_2^a\left[ m \right]} \right)}^{\rm{T}}}, \ldots ,{{\left( {{\bf{y}}_L^a\left[ m \right]} \right)}^{\rm{T}}}} \right]^{\rm{T}}}$ denotes the collective received signals of all AAUs at the CPU. Additionally, ${{\boldsymbol{\zeta }}_k}\left[ m \right] = {\left[ {{\boldsymbol{\zeta }}_{k,1}^{\rm{T}}\left[ m \right],{\boldsymbol{\zeta }}_{k,2}^{\rm{T}}\left[ m \right], \ldots ,{\boldsymbol{\zeta }}_{k,L}^{\rm{T}}\left[ m \right]} \right]^{\rm{T}}}$ denotes the collective data of UE $k$ that are errorneously processed.

\begin{figure*}[ht]
	\begin{small}
	\begin{equation}\label{y_l}
		\begin{aligned}
			{\bf{y}}_l^a\left[ m \right] 
			&= \sum\limits_{k = 1}^K {\sqrt {{p_k}} \left( {\left[ \begin{array}{c}
						\chi _{k,l,1}^m\frac{{W_{k,l,1}^{\left[ 0 \right]}}}{M}{{\bar h}_{k,l,1}}\left[ m \right]s_k^a\left[ m \right]\\
						\vdots \\
						\chi _{k,l,{N_{{\rm{RF}}}}}^m\frac{{W_{k,l,{N_{{\rm{RF}}}}}^{\left[ 0 \right]}}}{M}{{\bar h}_{k,l,{N_{{\rm{RF}}}}}}\left[ m \right]s_k^a\left[ m \right]
					\end{array} \right] + \underbrace {\sum\limits_{i \ne m}^{M - 1} {\left[ \begin{array}{c}
								\chi _{k,l,1}^i\frac{{W_{k,l,1}^{\left[ {{{\left( {m - i} \right)}_M}} \right]}}}{M}{{\bar h}_{k,l,1}}\left[ i \right]s_k^a\left[ i \right]\\
								\vdots \\
								\chi _{k,l,{N_{{\rm{RF}}}}}^i\frac{{W_{k,l,{N_{{\rm{RF}}}}}^{\left[ {{{\left( {m - i} \right)}_M}} \right]}}}{M}{{\bar h}_{k,l,{N_{{\rm{RF}}}}}}\left[ i \right]s_k^a\left[ i \right]
							\end{array} \right]} }_{{\rm{ICI}}} + \underbrace {{{\boldsymbol{\zeta }}_{k,l}}\left[ m \right]}_{{\rm{ISI}}}} \right)}  + {{\bf{n}}_l}\left[ m \right]\\
			&= \sum\limits_{k = 1}^K {\sqrt {{p_k}} \left( {\left[ \begin{array}{c}
						\kappa _{k,l,1,m}^m{{\bar h}_{k,l,1}}\left[ m \right]s_k^a\left[ m \right]\\
						\vdots \\
						\kappa _{k,l,{N_{{\rm{RF}}}},m}^m{{\bar h}_{k,l,{N_{{\rm{RF}}}}}}\left[ m \right]s_k^a\left[ m \right]
					\end{array} \right] + \underbrace {\sum\limits_{i \ne m}^{M - 1} {\left[ \begin{array}{c}
								\kappa _{k,l,1,m}^i{{\bar h}_{k,l,1}}\left[ i \right]s_k^a\left[ i \right]\\
								\vdots \\
								\kappa _{k,l,{N_{{\rm{RF}}}},m}^i{{\bar h}_{k,l,{N_{{\rm{RF}}}}}}\left[ i \right]s_k^a\left[ i \right]
							\end{array} \right]} }_{{\rm{ICI}}} + \underbrace {{{\boldsymbol{\zeta }}_{k,l}}\left[ m \right]}_{{\rm{ISI}}}} \right)}  + {{\bf{n}}_l}\left[ m \right]\\
			&= \sum\limits_{k = 1}^K {\sqrt {{p_k}} \left( {\Theta _{k,l,m}^m{{{\bf{\bar h}}}_{k,l}}\left[ m \right]s_k^a\left[ m \right] + \underbrace {\sum\limits_{i \ne m}^{M - 1} {\Theta _{k,l,m}^i{{{\bf{\bar h}}}_{k,l}}\left[ i \right]s_k^a\left[ i \right]} }_{{\rm{ICI}}} + \underbrace {{{\boldsymbol{\zeta }}_{k,l}}\left[ m \right]}_{{\rm{ISI}}}} \right)}  + {{\bf{n}}_l}\left[ m \right].
		\end{aligned}
	\end{equation}
	\end{small}%
	\hrulefill
\end{figure*}

\begin{figure*}[ht]
	\begin{equation}\label{y}
		{{\bf{y}}^a}\left[ m \right] = \sum\limits_{k = 1}^K {\sqrt {{p_k}} \left( {\Theta _{k,m}^m{{{\bf{\bar h}}}_k}\left[ m \right]s_k^a\left[ m \right] + \underbrace {\sum\limits_{i \ne m}^{M - 1} {\Theta _{k,m}^i{{{\bf{\bar h}}}_k}\left[ i \right]s_k^a\left[ i \right]} }_{{\rm{ICI}}} + \underbrace {{{\boldsymbol{\zeta }}_k}\left[ m \right]}_{{\rm{ISI}}}} \right)}  + {\bf{n}}\left[ m \right].
	\end{equation}
	\hrulefill
\end{figure*}

We utilize DU MR and MMSE combining vectors to detect the transmitted signal ${s_k^a\left[ m \right]}$ of UE $k$ in both centralized and distributed implementations. The expressions of combining vectors are consistent with those of downlink precoding vectors, without power normalization, which can refer to (\ref{w_mmse_c}) and (\ref{w_mmse_d}). The estimated signal of UE $k$ at the $m$-th subcarrier of the $a$-th OFDM symbol can be written as
\begin{equation}\label{s_k}
	\begin{aligned}
		\hat s_k^a\left[ m \right] &= {\bf{v}}_k^{\rm{H}}\left[ m \right]{{\bf{D}}_k}{{\bf{y}}^a}\left[ m \right]\\
		&= \overbrace {\sqrt {{p_k}} {\bf{v}}_k^{\rm{H}}\left[ m \right]{{\bf{D}}_k}\Theta _{k,m}^m{{{\bf{\bar h}}}_k}\left[ m \right]s_k^a\left[ m \right]}^{{\rm{Desired \; signal}}}\\
		&+ \overbrace {\sum\limits_{i = 1,i \ne k}^K {\sqrt {{p_i}} {\bf{v}}_k^{\rm{H}}\left[ m \right]{{\bf{D}}_k}\Theta _{i,m}^m{{{\bf{\bar h}}}_i}\left[ m \right]s_i^a\left[ m \right]} }^{{\rm{Inter \text{-} user \; interference}}}\\
		&+ \overbrace {\sum\limits_{i \ne m}^{M - 1} {\sum\limits_{j = 1}^K {\sqrt {{p_j}} {\bf{v}}_k^{\rm{H}}\left[ m \right]{{\bf{D}}_k}\Theta _{j,m}^i{{{\bf{\bar h}}}_j}\left[ i \right]s_j^a\left[ i \right]} } }^{{\rm{ICI}}}\\
		&+ \overbrace {\sum\limits_{i = 1}^K {\sqrt {{p_i}} {\bf{v}}_k^{\rm{H}}\left[ m \right]{{\bf{D}}_k}{{\boldsymbol{\zeta }}_i}\left[ m \right]} }^{{\rm{ISI}}} + \overbrace {{\bf{v}}_k^{\rm{H}}\left[ m \right]{{\bf{D}}_k}{\bf{n}}\left[ m \right]}^{{\rm{Noise}}},
	\end{aligned}
\end{equation}
where ${{\mathbf{v}}_k}\left[ m \right] = {\left[ {{\mathbf{v}}_{k,1}^{\text{T}}\left[ m \right],{\mathbf{v}}_{k,2}^{\text{T}}\left[ m \right], \ldots ,{\mathbf{v}}_{k,L}^{\text{T}}\left[ m \right]} \right]^{\text{T}}} \in {\mathbb{C}^{L{N_{{\text{RF}}}} \times 1}}$ denotes the collective combining vector of all AAUs to UE $k$ and ${\mathbf{n}}\left[ m \right] = {\left[ {{\mathbf{n}}_1^{\text{T}}\left[ m \right],{\mathbf{n}}_2^{\text{T}}\left[ m \right], \ldots ,{\mathbf{n}}_L^{\text{T}}\left[ m \right]} \right]^{\text{T}}} \in {\mathbb{C}^{L{N_{{\text{RF}}}} \times 1}}$ is the collective noise vector of all AAUs. Based on (\ref{s_k}), the uplink effective SINR of UE $k$ at the $m$-th subcarrier can be expressed as (\ref{SINR_k_asyn_U}).
\begin{figure*}[ht]
	\begin{small}
	\begin{equation}\label{SINR_k_asyn_U}
	\gamma _k^{{\text{Asyn-ul}}}\left[ m \right] = \frac{{{p_k}{{\left| {{\mathbf{v}}_k^{\text{H}}\left[ m \right]{{\mathbf{D}}_k}\Theta _{k,m}^m{{{\mathbf{\bar h}}}_k}\left[ m \right]} \right|}^2}}}{{\sum\limits_{i = 1,i \ne k}^K {{p_i}{{\left| {{\mathbf{v}}_k^{\text{H}}\left[ m \right]{{\mathbf{D}}_k}\Theta _{i,m}^m{{{\mathbf{\bar h}}}_i}\left[ m \right]} \right|}^2}}  + \sum\limits_{i \ne m}^{M - 1} {\sum\limits_{j = 1}^K {{p_j}{{\left| {{\mathbf{v}}_k^{\text{H}}\left[ m \right]{{\mathbf{D}}_k}\Theta _{j,m}^i{{{\mathbf{\bar h}}}_j}\left[ i \right]} \right|}^2}} }  + \sum\limits_{i = 1}^K {{p_i}{{\left| {{\mathbf{v}}_k^{\text{H}}\left[ m \right]{{\mathbf{D}}_k}{{\boldsymbol{\zeta }}_i}\left[ m \right]} \right|}^2}}  + \sigma _{{\text{ul}}}^2{{\left\| {{\mathbf{v}}_k^{\text{H}}\left[ m \right]{{\mathbf{D}}_k}} \right\|}^2}}}.
	\end{equation}
	\end{small}
	\hrulefill
\end{figure*}
Similar to the downlink scenario, for contrast analysis, the uplink effective SINR of UE $k$ at the $m$-th subcarrier in the hypothetical synchronous scenario can be expressed as
\begin{equation}\label{SINR_k_syn_U}
	\gamma _k^{{\text{Syn-ul}}}\left[ m \right] = \frac{{{p_k}{{\left| {{\mathbf{v}}_k^{\text{H}}\left[ m \right]{{\mathbf{D}}_k}{{{\mathbf{\bar h}}}_k}\left[ m \right]} \right|}^2}}}{{\sum\limits_{i = 1,i \ne k}^K {{p_k}{{\left| {{\mathbf{v}}_k^{\text{H}}\left[ m \right]{{\mathbf{D}}_k}{{{\mathbf{\bar h}}}_i}\left[ m \right]} \right|}^2}}  + \sigma _{{\text{ul}}}^2{{\left\| {{\mathbf{v}}_k^{\text{H}}\left[ m \right]{{\mathbf{D}}_k}} \right\|}^2}}}.
\end{equation}
In the small-cell scenario, the uplink SINR of UE $k$ served by AAU $l$ at the $m$-th subcarrier can be expressed as
\begin{equation}\label{SINR_kl_cellular_U}
	\gamma _{k,l}^{{\text{s}} \text{-} {\text{ul}}}\left[ m \right] = \frac{{{p_k}{{\left| {{\mathbf{v}}_{k,l}^{\text{H}}\left[ m \right]{{{\mathbf{\bar h}}}_{k,l}}\left[ m \right]} \right|}^2}}}{{\sum\limits_{i = 1,i \ne k}^K {{p_k}{{\left| {{\mathbf{v}}_{k,l}^{\text{H}}\left[ m \right]{{{\mathbf{\bar h}}}_{i,l}}\left[ m \right]} \right|}^2}}  + \sigma _{{\text{ul}}}^2{{\left\| {{\mathbf{v}}_{k,l}^{\text{H}}\left[ m \right]} \right\|}^2}}}.
\end{equation}
The achievable uplink SE of UE $k$ at the $m$-th subcarrier in the cell-free and small-cell scenarios are given by (\ref{SE_k}) and (\ref{SE_kl_cellular}), respectively.

\section{Joint Beam selection and UE association} \label{Beam selection and UE association}
In this section, we formulate the problem of maximizing the sum rate by incorporating various aspects such as beam selection, AAU-UE association, constraints on the number of RF chains and beam conflict control (BCC). We propose two different algorithms primarily utilize beam-domain channel amplitudes and large-scale fading coefficients, respectively.
\subsection{Problem Formulation}
The mmWave channel matrix in the beam domain is sparse. The transmitter sends a beam with a high gain corresponding to the channel vector of the UE, ensuring approximately optimal communication quality. With the limited number of RF chains, ${N_{{\rm{RF}}}} \le N$ , each AAU can only select a maximum of ${N_{{\rm{RF}}}}$ orthogonal beams to maximize array gain. Additionally, each beam is dedicated to serving a single UE, meaning the BCC constraint must be satisfied \cite{sun2019beam}. Given that ${N_{{\rm{RF}}}} \le K$, the number of UEs each AAU can associate with is limited. Hence, an AAU-UE association process is necessary. The joint beam selection and UE association problem can be formulated as the following optimization task:
\begin{subequations}
	\begin{align}
		\mathop {{\rm{maximize}}}\limits_{\left\{ {{b_{k,l}},{u_{k,l}}:k \in {{\cal K}},l \in {{\cal L}}} \right\}} &\sum\limits_{i = 1}^K {{R_i}\left[ m \right]} \label{opt}\\
		\text{subject to} ~~
		&{b_{k,l}} \ne {b_{k',l}},~\forall k \ne k', \label{opt:sub1}\\
		&{b_{k,l}} \in {{{\cal B}}_l},~\forall k \in {{\cal K}},l \in {{\cal L}},\label{opt:sub2}\\
		&\left| {{\Gamma _l}} \right| \le {N_{{\rm{RF}}}},~\forall l \in {{\cal L}}, \label{opt:sub3}\\
		&\sum\limits_{i = 1}^K {{u_{i,l}}}  \le {N_{{\rm{RF}}}},~\forall l \in {{\cal L}}, \label{opt:sub4}
	\end{align}
\end{subequations}
where ${\cal K} \buildrel \Delta \over = \left\{ {1, \ldots ,K} \right\}$ and ${\cal L} \buildrel \Delta \over = \left\{ {1, \ldots ,L} \right\}$ are defined as the index sets of UEs and AAUs, respectively. ${b_{k,l}}$ is the index of the beam assigned by AAU $l$ to UE $k$. If UE $k$ is not served by AAU $l$, meaning AAU $l$ does not allocate any beams to UE $k$, we set ${b_{k,l}} = 0$. ${\mathcal{B}_l}$ is the set of indexes of selectable beams of AAU $l$. ${\Gamma _l}$ is the set of indexes of the beams selected by AAU $l$. ${u_{k,l}}$ stands for the association indicator between AAU $l$ and UE $k$.

The first constraint (\ref{opt:sub1}) enforces the BCC constraint, ensuring that different UEs cannot select the same beams of AAU $l$. The second constraint (\ref{opt:sub2}) implies that each beam direction is chosen from a predefined beam codebook, consistent with the characteristics of the DLA. The third constraint (\ref{opt:sub3}) restricts the maximum number of beam directions selected by each AAU to no more than ${N_{{\rm{RF}}}}$. Finally, the last constraint (\ref{opt:sub4}) indicates that the number of UEs associated with each AAU does not exceed ${N_{{\text{RF}}}}$.

The optimization problem presented above is in general challenging, due to the complexity of the objective function $\sum\limits_{i = 1}^K {{R_i}\left[ m \right]}$ and the coupling properties of the beam selection and UE association problems. In particular, for the considered AAUs equipped with a large number of antennas (beams) in the mmWave cell-free massive MIMO system, the optimal solution has to be found through an exhaustive search. To obtain feasible solutions with relatively low complexity, we propose two suboptimal strategies for beam selection and UE association. They primarily rely on beam-domain channel amplitudes and large-scale fading coefficients, respectively. Additionally, We will also consider the impacts of asynchronous reception to enhance the adaptability of the proposed strategies in asynchronous scenarios.

\subsection{Based on Beam-domain Channel Amplitude}
The first strategy primarily relies on beam-domain CSIs and is divided into two stages. In the first stage, each AAU selects the best beam for each UE based on the maximum magnitude selection (MM-S) strategy for beam-domain channels, which can be expressed as
\begin{equation}\label{b_kl}
	{b_{k,l}} = \mathop {\arg \max }\limits_{b \in {\mathcal{B}_l}} {\left| {{{{\mathbf{\tilde h}}}_{k,l}}\left( b \right)} \right|^2},
\end{equation}
where ${{\mathbf{\tilde h}}_{k,l}}\left( b \right)$ is the $b$-th element of ${{\mathbf{\tilde h}}_{k,l}}$. To fulfill BCC, the search is performed sequentially, commencing from UE ${k^ * }$ with the lowest large-scale fading. This ensures that the UEs with superior channel conditions can be allocated better beams, as follows:
\begin{equation}\label{k_search}
	{k^ * } = \mathop {\arg \min }\limits_{k' \in \mathcal{{\cal K}}} {\left| {\left( {\frac{M}{{M - {\varepsilon _{k',l}}}}} \right){\beta _{k',l}}} \right|^2},
\end{equation}
where the large-scale fading coefficient ${\beta _{k',l}}$ is attached by a gain factor ${\left( {\frac{M}{{M - {\varepsilon _{k',l}}}}} \right)}$ to reflect the impact of asynchronous reception.

In the second stage, following the selected optimal beam for each UE, the ${N_{{\rm{RF}}}}$ UEs with the highest channel gain are chosen for association, that is,
\begin{equation}\label{k}
	k = \mathop {\arg \max }\limits_{k' \in \mathcal{{\cal K}}} {\left| {\left( {\frac{{M - {\varepsilon _{k',l}}}}{M}} \right){{{\bf{\tilde h}}}_{k',l}}\left( {{b_{k'l}}} \right)} \right|^2},
\end{equation}
where the optimal beam-domain channel ${{{{\bf{\tilde h}}}_{k',l}}\left( {{b_{k'l}}} \right)}$ is attached by a discount coefficient ${\left( {\frac{{M - {\varepsilon _{k',l}}}}{M}} \right)}$ to incorporate the influence of asynchronous timing offsets exceeding the CP range.

The computational complexity of the algorithm is ${{\cal O}}\left( {L\left( {K + {N_{{\rm{RF}}}}} \right)} \right)$. The detailed algorithm dominated by beam-domain channel amplitudes for addressing (\ref{opt}) is summarized in Algorithm 1.

\begin{algorithm}[!ht]\label{algorithm1}
	\renewcommand{\algorithmicrequire}{\textbf{Input:}}
	\renewcommand{\algorithmicensure}{\textbf{Output:}}
	\caption{Proposed beam selection and UE association algorithm mainly based on beam-domain channel amplitude.}
	\begin{algorithmic}[1] 
		\REQUIRE $\left\{ {{\beta _{k,l}},k \in {{\cal K}},l \in {{\cal L}}} \right\}$ and $\left\{ {{{{\bf{\tilde h}}}_{k,l}},k \in {{\cal K}},l \in {{\cal L}}} \right\}$; 
		\ENSURE $\left\{ {{b_{k,l}},{u_{k,l}}:k \in {{\cal K}},l \in {{\cal L}}} \right\}$; 
		
		\STATE Initialize $l = 1;$
		\WHILE {$l \le L$}
			\STATE Set ${{{\cal B}}_l} = \left\{ {1,2, \ldots ,N} \right\}$ and ${{\cal K}} = \left\{ {1,2, \ldots ,K} \right\}$;
			\WHILE {${{\cal K}} \ne \emptyset $}
				\STATE Search for the UE ${k^ * }$ with the smallest large-scale using (\ref{k_search}).
				\STATE Update ${{\cal K}} \leftarrow {{\cal K}}\backslash \left\{ {{k^ * }} \right\}$.
				\STATE Obtain the best beam ${b_{{k^ * },l}}$ assigned by AAU $l$ to UE $k^*$ using (\ref{b_kl}).
				\STATE Update ${{\mathcal{B}}_l} \leftarrow {{\mathcal{B}}_l}\backslash {b_{{k^ * },l}}$.
			\ENDWHILE
			\STATE Set $n = 1$, ${{\cal K}} = \left\{ {1,2, \ldots ,K} \right\}$ and ${{{\cal D}}_l} = \emptyset$.
			\WHILE {$n \le {N_{{\rm{RF}}}}$}
				\STATE Select the UE $k$ with largest magnitude of optimal beam using (\ref{k}).
				\STATE Associate AAU $l$ and UE $k$, i.e., set ${u_{k,l}} = 1$ and ${{{\cal D}}_l} \leftarrow {{{\cal D}}_l} \cup \left\{ k \right\}$.
				\STATE Update $n \leftarrow n + 1$ and ${{\cal K}} \leftarrow {{\cal K}}\backslash \left\{ k \right\}$.
			\ENDWHILE
			\STATE Update $l \leftarrow l + 1$.
		\ENDWHILE
	\end{algorithmic}
\end{algorithm}

\subsection{Based on Large-scale Fading Coefficient}
The second strategy conducts AAU-UE association according to the large-scale fading coefficient, thus eliminating the need to reassociate UEs for the slight changes in CSIs. Instead, it simply needs to adjust the beam assigned to each UE based on the channel magnitudes. The overall computational overhead in resource scheduling is reduced.

The second strategy is still divided into two stages. First, each AAU is associated with ${N_{{\rm{RF}}}}$ UEs that have the smallest large-scale fading. Next, an optimal beam  is selected for each associated UE in turn, based on the large-scale fading of the UEs in ascending order. This process needs to ensure that the BCC constraint is satisfied. The computational complexity of the algorithm is ${{\cal O}}\left( {L{N_{{\rm{RF}}}}} \right)$. The detailed algorithm dominated by large-scale fading coefficient is summarized in Algorithm 2.

\begin{algorithm}[!ht]\label{algorithm2}
	\renewcommand{\algorithmicrequire}{\textbf{Input:}}
	\renewcommand{\algorithmicensure}{\textbf{Output:}}
	\caption{Proposed beam selection and UE association algorithm mainly based on large-scale fading coefficient.}
	\begin{algorithmic}[1] 
		\REQUIRE $\left\{ {{\beta _{k,l}},k \in {{\cal K}},l \in {{\cal L}}} \right\}$ and $\left\{ {{{{\bf{\tilde h}}}_{k,l}},k \in {{\cal K}},l \in {{\cal L}}} \right\}$; 
		\ENSURE $\left\{ {{b_{k,l}},{u_{k,l}}:k \in {{\cal K}},l \in {{\cal L}}} \right\}$; 
		
		\STATE Initialize $l = 1;$
		\WHILE {$l \le L$}
			\STATE Set $n = 1$, ${{{\cal D}}_l} = \emptyset $, ${{{\cal B}}_l} = \left\{ {1,2, \ldots ,N} \right\}$ and ${{\cal K}} = \left\{ {1,2, \ldots ,K} \right\}$.
			\WHILE {$n \le {N_{{\rm{RF}}}}$}
				\STATE Select the UE ${k^ * }$ with the smallest large-scale fading coefficient using (\ref{k_search}).
				\STATE Associate AAU $l$ and UE ${k^ * }$, i.e., set ${u_{{k^ * },l}} = 1$.
				\STATE Update ${{{\cal D}}_l} \leftarrow {{{\cal D}}_l} \cup \left\{ {{k^ * }} \right\}$ and ${{\cal K}} \leftarrow {{\cal K}}\backslash \left\{ {{k^ * }} \right\}$.
				\STATE AAU $l$ assigns optimal beam ${b_{{k^ * },l}}$ to UE ${k^ * }$ based on the maximum magnitude strategy using (\ref{b_kl}).
				\STATE Update $n \leftarrow n + 1$ and ${{\mathcal{B}}_l} \leftarrow {{\mathcal{B}}_l}\backslash {b_{{k^ * },l}}$.
			\ENDWHILE
			\STATE Update $l \leftarrow l + 1$.
		\ENDWHILE
	\end{algorithmic}
\end{algorithm}

\section{Simulation Results} \label{Simulation Results}
In this section, we provide simulation results to demonstrate the performance of cell-free mmWave massive MIMO-OFDM systems with our proposed hybrid precoding architecture based on the PBTA scheme. We consider a simulation scenario, where $L=30$ AAUs and $K=20$ UEs are individually and uniformly distributed in a square area of $2 \times 2$ $\rm{km}^2$, and a wrap-around technique is employed. The AAUs are deployed 10 meters above the UEs, ensuring a minimum distance. Each AAU is equipped with $N = 50$ antennas and ${N_{{\rm{RF}}}} = 8$ RF chains. The uplink transmission power of each UE is ${p_k} = 100{\rm{mW}}$ and the downlink transmission power of each AAU is ${\rho _{\max }} = 4{\rm{W}}$. The typical mmWave carrier frequency, 28GHz, is considered \cite{you2017bdma}. The system bandwidth is $B = 100{\rm{MHz}}$ and subcarrier space is $\Delta f = 120{\rm{kHz}}$. There are a total of $M = 128$ subcarriers. In order to fully reflect the impacts caused by asynchronous reception, the sampling length of CP is ${M_{{\rm{CP}}}} = 10$ and the channel delay spread is ${T_{\rm{D}}} = 3$. The path loss exponent is $\vartheta=2$ and the standard deviation $\varsigma $ of the shadow fading factor ${A_\varsigma }$ is 4 dB. The complex gain, delay and departure angle of the $p$-th path are modeled as $\alpha _{k,l}^p \sim {{\cal C}{\cal N}}\left( {0,1} \right)$, ${\tau _p} \sim {{\cal U}}\left( {0,200{\rm{ns}}} \right)$ and ${\psi _{k,l,p}} \sim {{\cal U}}\left( {{{ - \pi } \mathord{\left/{\vphantom {{ - \pi } 2}} \right.\kern-\nulldelimiterspace} 2},{\pi  \mathord{\left/{\vphantom {\pi  2}} \right.\kern-\nulldelimiterspace} 2}} \right)$ respectively. Noise figure is ${\rm{NF}} = 9{\rm{dB}}$, and the noise power is ${\sigma ^2} =  - 174{\rm{dBm/Hz}} + 10{\log _{10}}\left( B \right) + {\rm{NF}}$. Since we use DU precoding, which will eliminate the impact of phase shift, the different asynchronous phase offsets caused by different subcarrier indexes have little impact on our subsequent simulations.

\begin{figure}[!t]
	\centering
	\includegraphics[scale=0.5]{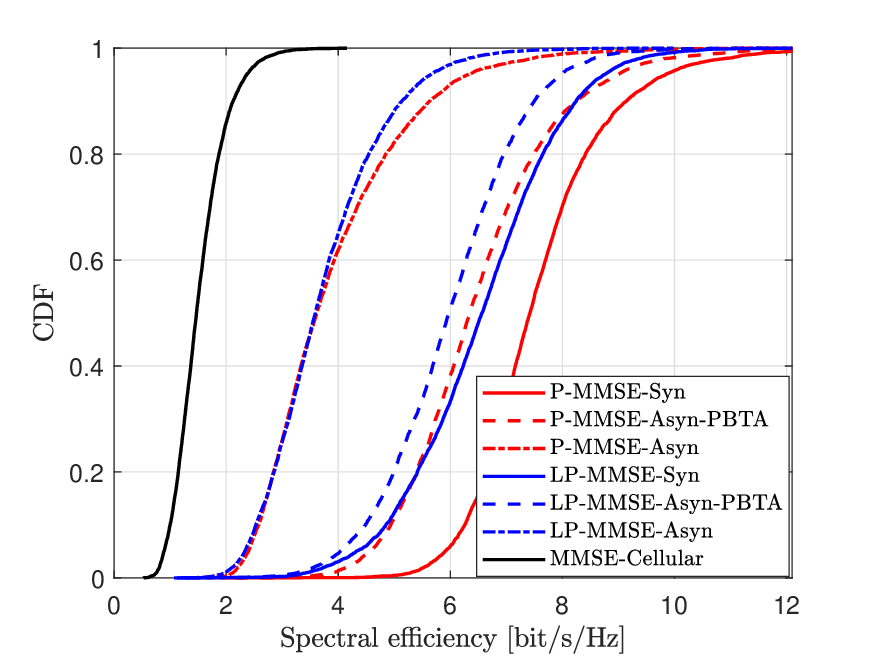}
	\caption{The CDF of downlink SE comparison of synchronous (Syn), asynchronous (Asyn), PBTA and small-cell (Cellular) scenarios when using centralized P-MMSE and distributed LP-MMSE precoding, respectively.}
	\label{figure5}
\end{figure}
\begin{figure}[!t]
	\centering
	\includegraphics[scale=0.5]{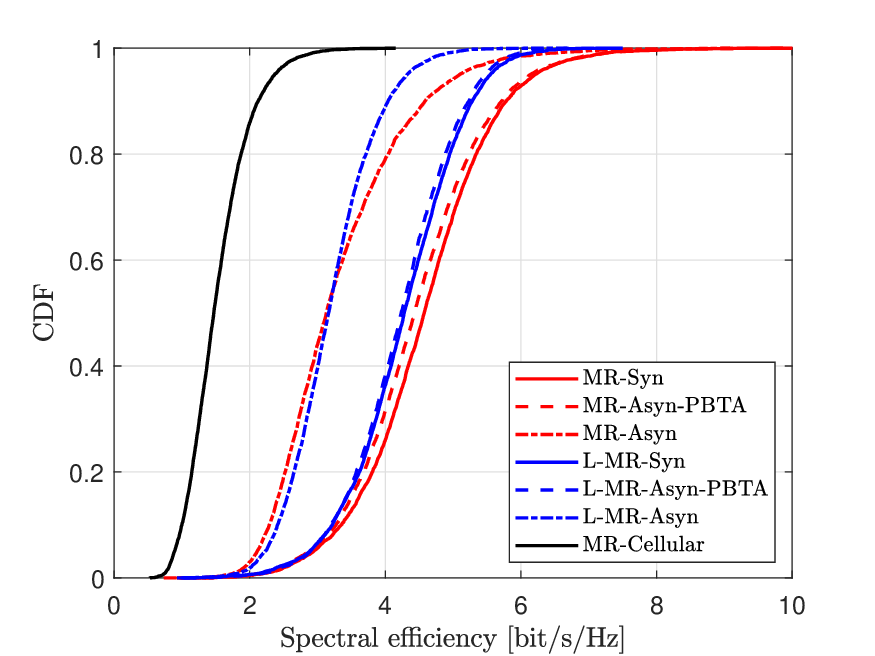}
	\caption{The CDF of downlink SE comparison of synchronous (Syn), asynchronous (Asyn), PBTA and small-cell (Cellular) scenarios when using centralized and distributed MR precoding, respectively.}
	\label{figure6}
\end{figure}

In Fig. \ref{figure5} and Fig. \ref{figure6}, we compare the downlink performances across synchronous, asynchronous, the proposed PBTA and small-cell scenarios\footnote{In asynchronous scenarios, a nearest UE-based timing-advance scheme is utilized for comparison \cite{chowdhury2023resilient}, whereas in PBTA scenarios, our proposed PBTA architecture is employed.}, comparing centralized P-MMSE precoding, distributed LP-MMSE precoding, centralized MR precoding and distributed L-MR precoding, respectively. The key observations are made as follows: Firstly, both centralized and distributed implementations experience significant performance degradation in asynchronous scenarios compared to synchronous ones. This is primarily due to the interference caused by ICI and ISI resulting from the asynchronous timing offset exceeding the CP range. However, asynchronous scenarios still outperform small-cell network scenarios. Secondly, the implementation of our proposed PBTA results in substantial performance improvements. PBTA scheme ensures that the desired signals arrive simultaneously, thereby eliminating a large portion of the interference generated by asynchronous reception. This indicates that our proposed PBTA can effectively mitigate asynchronous interference. Thirdly, the performance loss due to asynchronous reception is more severe in centralized processing than in distributed processing. After applying PBTA, the performance gain is greater for distributed processing, bringing it closer to synchronous performance than centralized processing. This indicates that centralized implementations are much more sensitive to asynchronous interference than distributed implementations. Fourthly, in asynchronous scenarios, the performance loss with MR precoding is relatively less than that with MMSE precoding. After implementing PBTA, the performance of MR precoding is closer to that of synchronous scenarios. This suggests that MMSE precoding is much more sensitive to asynchronous reception than MR precoding. Fifthly, there remains a performance gap between the PBTA and the synchronous scenarios. This is mainly due to the ICI and the ISI caused by the asynchronous reception of interference signals. While the ICI of the desired signals has been eliminated, the asynchronous timing offset of some interference signals still falls outside the CP range, resulting in a small portion of the ICI and ISI still existing.

\begin{figure}[!t]
	\centering
	\includegraphics[scale=0.5]{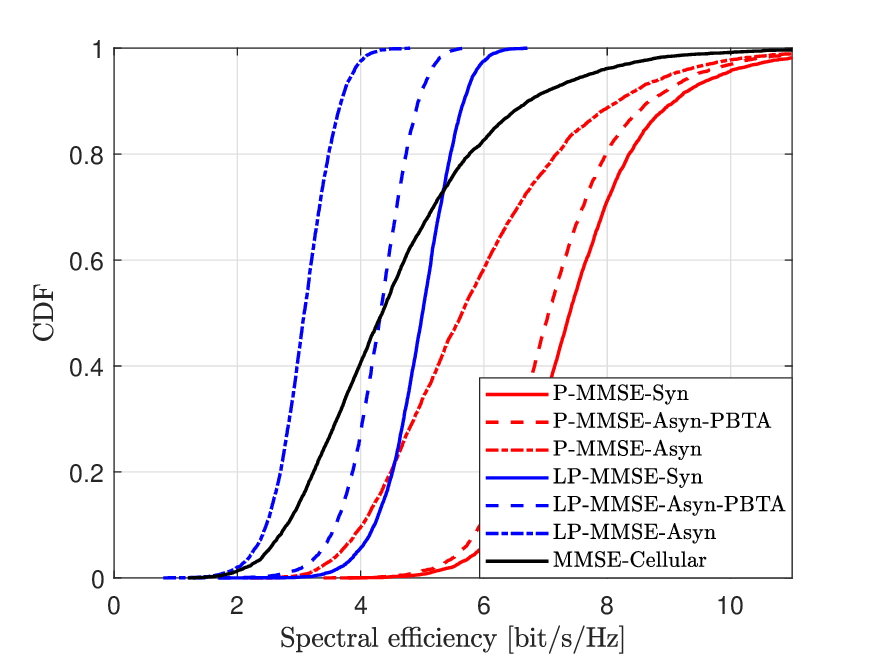}
	\caption{The CDF of uplink SE comparison of synchronous (Syn), asynchronous (Asyn), PBTA and small-cell (Cellular) scenarios when using centralized P-MMSE and distributed LP-MMSE combining, respectively.}
	\label{figure7}
\end{figure}
\begin{figure}[!t]
	\centering
	\includegraphics[scale=0.5]{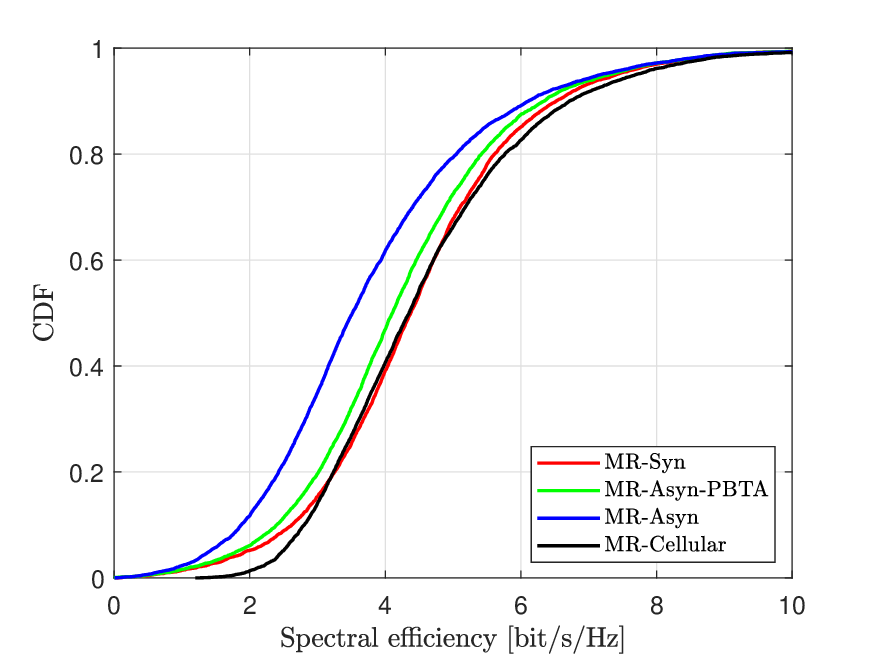}
	\caption{The CDF of uplink SE comparison of synchronous (Syn), asynchronous (Asyn), PBTA and small-cell (Cellular) when using MR combining.}
	\label{figure8}
\end{figure}

In Fig. \ref{figure7} and Fig. \ref{figure8}, we compare the uplink performances across synchronous, asynchronous, the proposed PBTA and small-cell scenarios, comparing centralized P-MMSE combining, distributed LP-MMSE combining and MR combining, respectively. In the uplink, asynchronous scenarios have a large performance degradation for both centralized and distributed MMSE combining compared to synchronous scenarios. The PBTA scheme can closely approximate the performance of synchronous scenarios. It indicates that our proposed PBTA is also effective in uplink scenarios. The uplink performance of distributed MR combining is consistent with that of the centralized implementation. Similar to the downlink results, the performance loss with MR combining is less than that with MMSE combining in asynchronous scenarios. After implementing PBTA, the performance is closer to synchronization. This also indicates that the MMSE combining is much more sensitive to asynchronous reception than the MR combining. 

\begin{figure}[!t]
	\centering
	\includegraphics[scale=0.5]{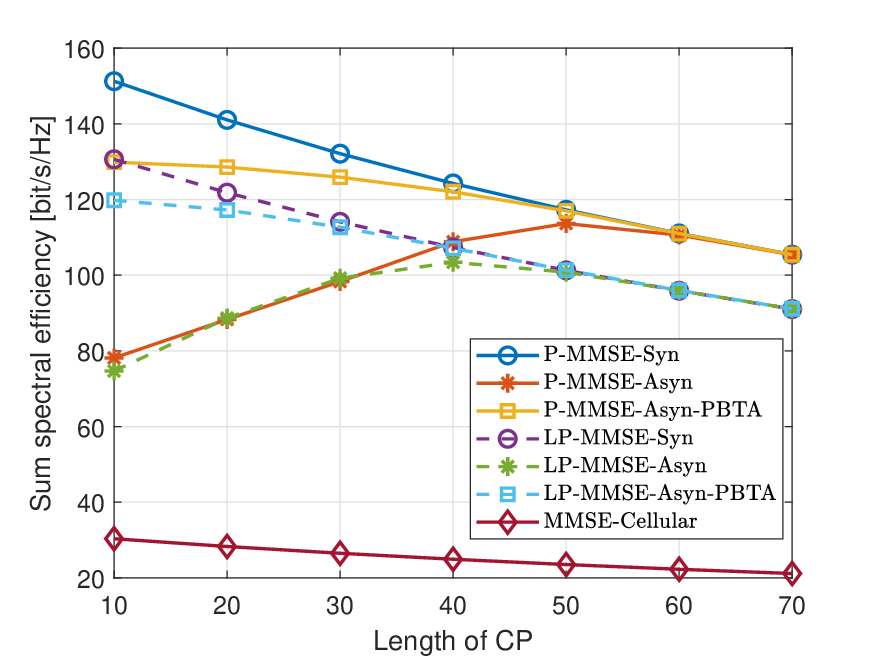}
	\caption{The downlink sum SE comparison versus the length of CP.}
	\label{figure9}
\end{figure}

Fig. \ref{figure9} presents the downlink sum SE against the length of CP. We adopt the same simulation parameters as in Fig. \ref{figure5} except for the varying length of CP. The length of CP is set to $10 \sim 70$. The key observations are made as follows: Firstly, as the CP length increases, the sum SE gradually decreases for synchronous, PBTA, and small-cell scenarios. It is due to the fact that as the CP length increases, the percentage of effectively transmitted data decreases, leading to a decrease in overall performance, i.e., the pre-log factor ${\frac{M}{{M + {M_{{\rm{CP}}}}}}}$ decreases as ${M_{{\rm{CP}}}}$ increases. Secondly, in the PBTA scenarios, when the CP length is small, the degradation in SE is less pronounced than the synchronous scenario. This is because an increase in CP length at this time is conducive to the elimination of the residual asynchronous interference, resulting in performance gains. As the CP length continues to increase, the asynchronous delay can be completely contained within the CP range, leading to similar performance degradation as the synchronous scenarios. Thirdly, for asynchronous scenarios, increasing the CP length initially causes the SE to gradually increase to a peak before it starts to decay. This is due to that the asynchronous interference is very severe when the CP length is short, dominating most of the performance loss. In this case, increasing the CP length significantly improves performance. However, with the CP length increased to a certain degree, the CP length is sufficiently long to encompass the asynchronous timing offset. At this time, the decrease of the pre-log factor ${\frac{M}{{M + {M_{{\rm{CP}}}}}}}$ will become the dominant factor in SE attenuation, causing the SE to gradually decrease, which is consistent with the trend of the synchronous scenario.

\begin{figure}[!t]
	\centering
	\includegraphics[scale=0.5]{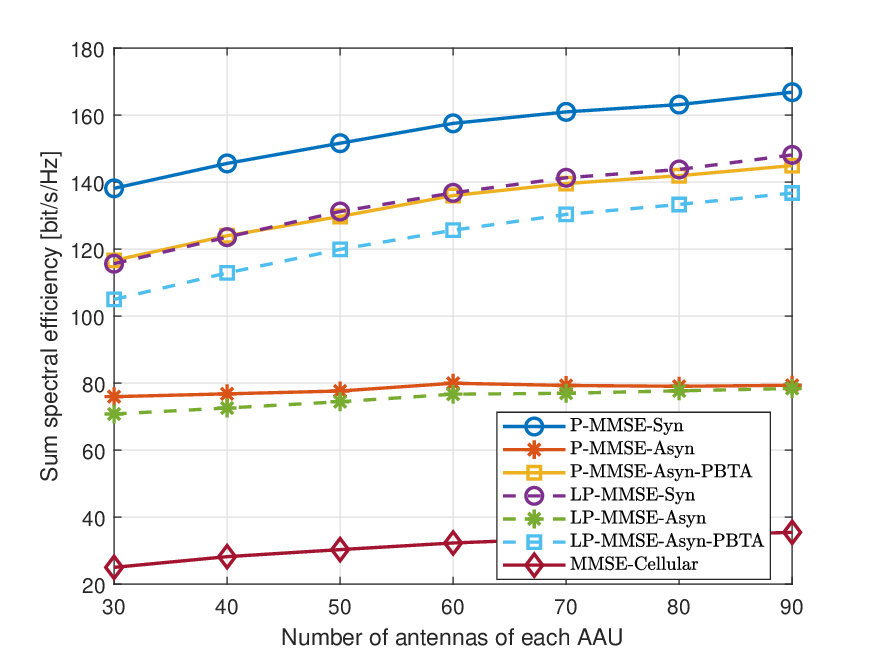}
	\caption{The downlink sum SE comparison versus the number of antennas of each AAU.}
	\label{figure10}
\end{figure}

In Fig. \ref{figure10}, we present the downlink sum SE against the number of antennas of each AAU. We adopt the same simulation parameters as in Fig. \ref{figure5} except for the varying number of antennas of each AAU. The increase in the number of antennas means that more precise beam directions can be selected, enabling the formation of narrower beams and enhancing inter-UE interference cancellation. Fig. \ref{figure10} shows that for synchronous, PBTA, and small-cell scenarios, the SE gradually improves as the number of antennas increases. This improvement reflects the enhanced interference cancellation capability of narrower beams. However, for asynchronous scenarios, ICI and ISI caused by asynchronous timing offsets are the primary factors affecting performance. It is observed that merely increasing the number of antennas to get narrower beams cannot effectively improve the SE in the asynchronous scenarios. In contrast, by generating narrower beams, our proposed PBTA scheme achieves better interference cancellation and thus improves SE. This indicates that PBTA effectively eliminates most asynchronous interference, and the residual asynchronous interference can be further suppressed by increasing the antenna array size and generating narrower beams, thereby enhancing overall performance. 

\begin{figure}[!t]
	\centering
	\includegraphics[scale=0.5]{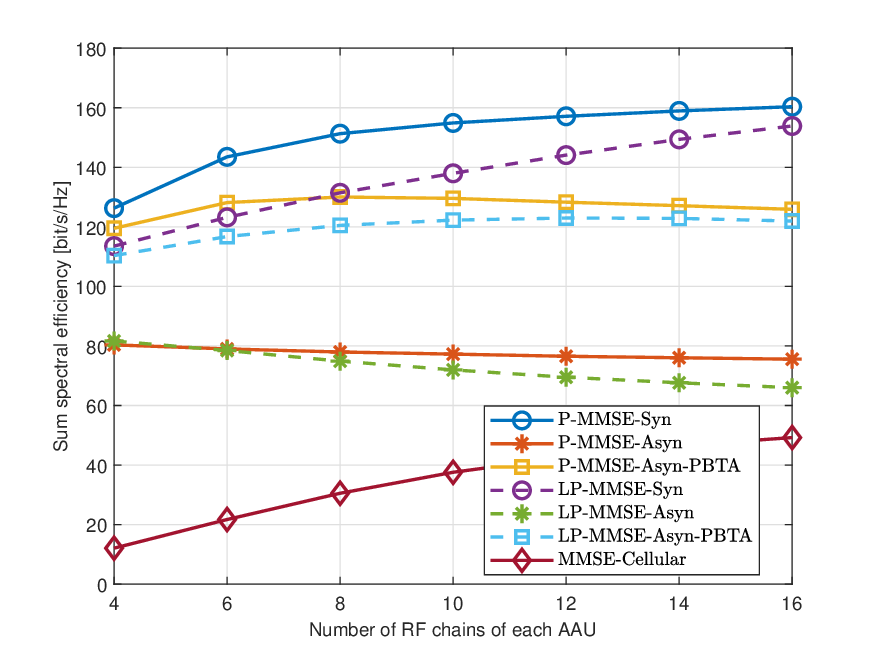}
	\caption{The downlink sum SE comparison versus the number of RF chains of each AAU.}
	\label{figure11}
\end{figure}

In Fig. \ref{figure11}, we present the downlink sum SE against the number of RF chains of each AAU. We adopt the same simulation parameters as in Fig. \ref{figure5} except for the varying number of RF chains of each AAU. The increase in the number of RF chains allows each AAU to select more beams and associate with more UEs, but it also increases inter-UE interference. Fig. \ref{figure11} shows that for synchronous and small-cell scenarios, the SE is gradually improved with the increase of the number of RF chains. This improvement is due to the strong inter-UE interference cancellation ability in synchronous scenarios, ensuring a stable enhancement in SE. In small-cell scenarios, the increase in RF chains allows each AAU to select more beams, making it more likely for each UE to connect with an AAU that has better channel conditions. For the PBTA scenario, the increase in RF chains provides limited improvement in SE and may even lead to degradation. This is because PBTA retains some asynchronous interference caused by inter-UE interference. As the number of RF chains increases, so does the inter-UE interference, resulting in greater asynchronous interference. In addition, for asynchronous scenarios, increasing the number of RF chains actually worsens the SE. This indicates that each AAU associating with more UEs has little impact on the desired signal power but introduces stronger interference due to the ICI and ISI caused by asynchronous reception.

\begin{figure}[!t]
	\centering
	\includegraphics[scale=0.5]{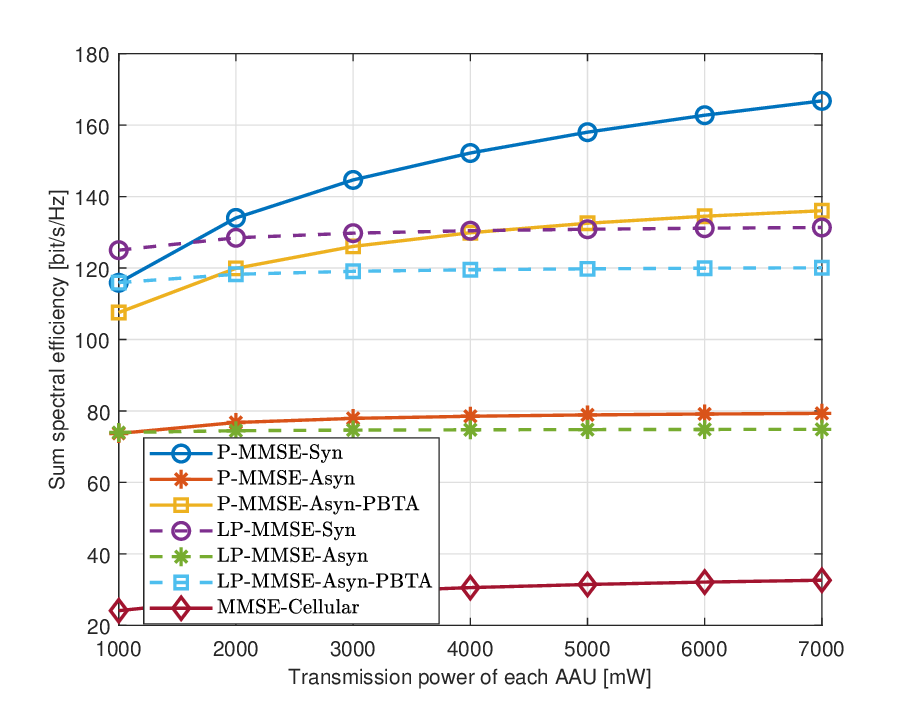}
	\caption{The downlink sum SE comparison versus the transmission power of each AAU.}
	\label{figure12}
\end{figure}

In Fig. \ref{figure12}, we present the downlink sum SE against the transmission power of each AAU. We adopt the same simulation parameters as in Fig. \ref{figure5} except for the varying transmission power of each AAU. It can be seen from the figure that with the increase of the transmission power, the SE of the synchronous and PBTA scenarios is gradually improved in the centralized implementation. The improvement of SE in the distributed implementation is minimal. Boosting the transmission power has little impact on the performance of asynchronous scenarios, indicating that it is not possible to improve asynchronous reception by means of power boosting. Our proposed PBTA scheme has a significant improvement for asynchronous reception and is able to further improve performance by boosting the transmission power.

\begin{figure}[!t]
	\centering
	\includegraphics[scale=0.5]{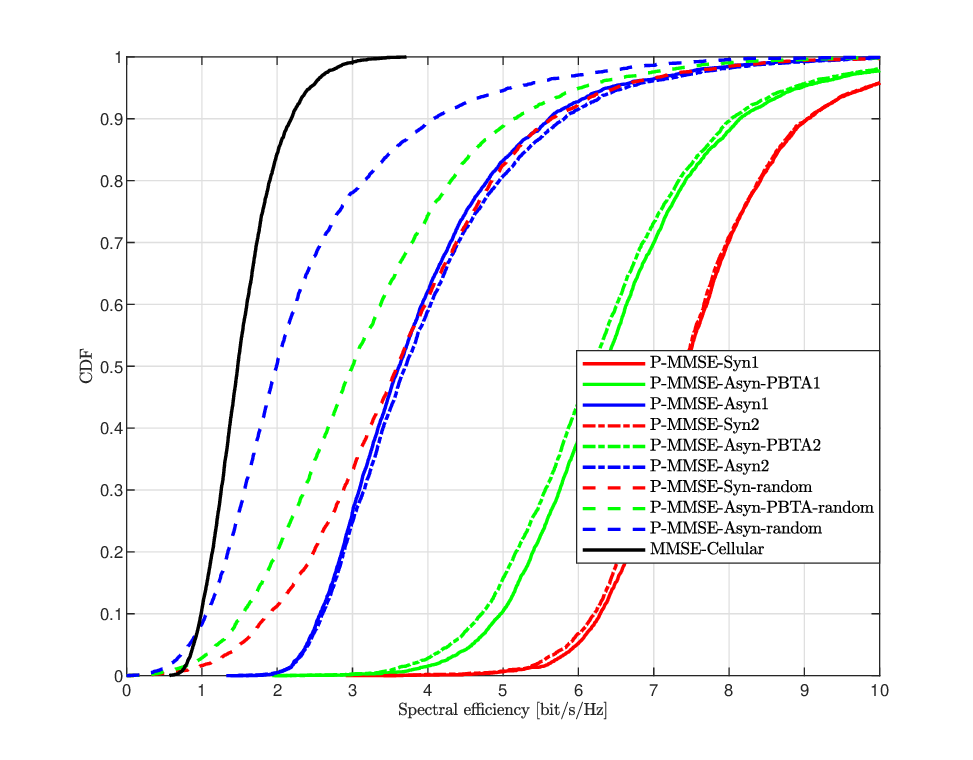}
	\caption{The CDF of centralized downlink SE comparison of synchronous (Syn), asynchronous (Asyn), PBTA and small-cell (Cellular) scenarios with diffenent beam selection and UE association algorithms.}
	\label{figure13}
\end{figure}
\begin{figure}[!t]
	\centering
	\includegraphics[scale=0.5]{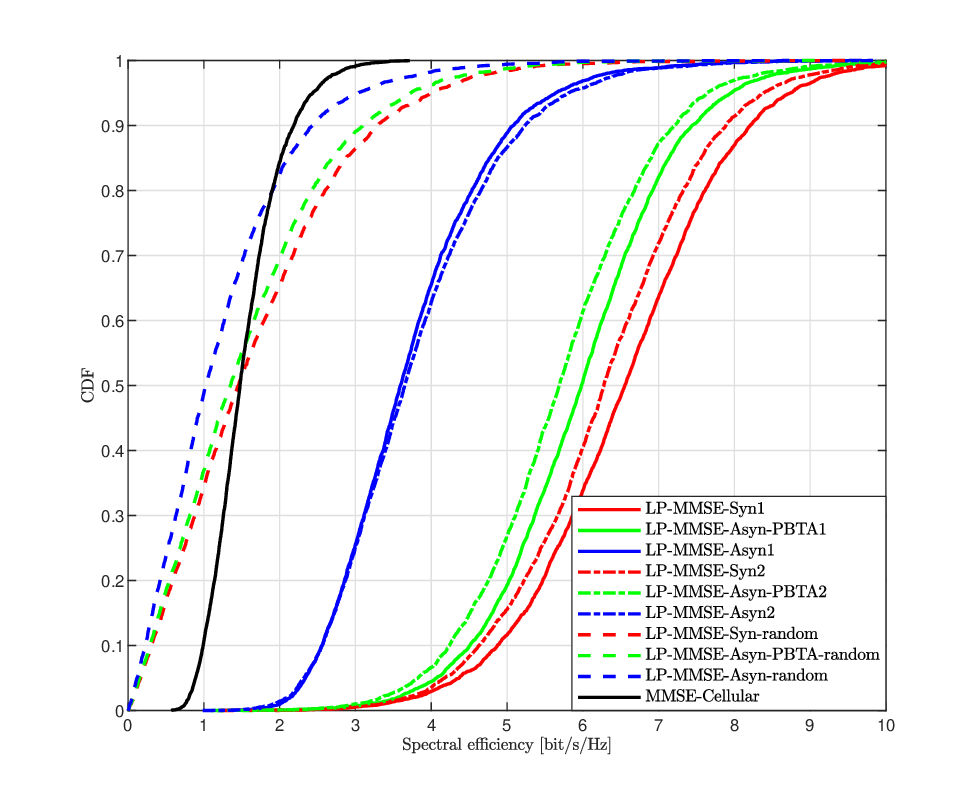}
	\caption{The CDF of distributed downlink SE comparison of synchronous (Syn), asynchronous (Asyn), PBTA and small-cell (Cellular) scenarios with diffenent beam selection and UE association algorithms.}
	\label{figure14}
\end{figure}

In Fig. \ref{figure13} and Fig. \ref{figure14}, we compare the downlink performance of synchronous, asynchronous, the proposed PBTA and small-cell scenarios with different joint beam selection and UE association algorithms, comparing centralized P-MMSE precoding and distributed LP-MMSE precoding, respectively. The legend suffix "1" indicates the use of Algorithm 1, "2" indicates the use of Algorithm 2 and "random" indicates the random selection of beams and UE associations. We can observe that compared to random selection, our proposed two suboptimal low-complexed algorithms show significant performance improvements, indicating their effectiveness. In addition, Algorithm 1 yields better performance in synchronous and PBTA scenarios due to its more refined approach. However, the computational overhead and complexity are higher. Algorithm 2 provides better performance in the asynchronous scenarios. Its resistance to asynchronous interference is superior to that of Algorithm 1.

\section{Conclusion} \label{Conclusion}
In this paper, we presented a comprehensive asynchronous beam-domain signal transmission model for mmWave CF mMIMO-OFDM systems in both downlink and uplink. We emphasized that ICI and ISI resulting from asynchronous reception can significantly reduce the achievable rate in the wide area coverage scenarios. To solve this issue, we proposed a novel PBTA hybrid precoding architecture and derived the SE for downlink and uplink asynchronous receptions, considering phase shift, ICI and ISI. In addition, we developed two suboptimal low-complexity joint beam selection and UE association algorithms for the asynchronous scenarios, which considered the impact of asynchronous timing offset exceeding the CP range. 

Simulation results demonstrated that our proposed PBTA can effectively mitigate asynchronous interference compared to the nearest AAU/UE-based timing-advance scheme. Furthermore, our proposed joint beam selection and UE association algorithms can improve SE performance effectively with low complexity.

\bibliography{ref}

\begin{thebibliography}{10}
\providecommand{\url}[1]{#1}
\csname url@samestyle\endcsname
\providecommand{\newblock}{\relax}
\providecommand{\bibinfo}[2]{#2}
\providecommand{\BIBentrySTDinterwordspacing}{\spaceskip=0pt\relax}
\providecommand{\BIBentryALTinterwordstretchfactor}{4}
\providecommand{\BIBentryALTinterwordspacing}{\spaceskip=\fontdimen2\font plus
\BIBentryALTinterwordstretchfactor\fontdimen3\font minus
  \fontdimen4\font\relax}
\providecommand{\BIBforeignlanguage}[2]{{%
\expandafter\ifx\csname l@#1\endcsname\relax
\typeout{** WARNING: IEEEtran.bst: No hyphenation pattern has been}%
\typeout{** loaded for the language `#1'. Using the pattern for}%
\typeout{** the default language instead.}%
\else
\language=\csname l@#1\endcsname
\fi
#2}}
\providecommand{\BIBdecl}{\relax}
\BIBdecl

\bibitem{you2021towards}
X.~You, C.-X. Wang, J.~Huang, X.~Gao, Z.~Zhang, M.~Wang, Y.~Huang, C.~Zhang,
  Y.~Jiang, J.~Wang \emph{et~al.}, ``Towards 6{G} wireless communication
  networks: Vision, enabling technologies, and new paradigm shifts,''
  \emph{Science China Information Sciences}, vol.~64, pp. 1--74, 2021.

\bibitem{ammar2021user}
H.~A. Ammar, R.~Adve, S.~Shahbazpanahi, G.~Boudreau, and K.~V. Srinivas,
  ``User-centric cell-free massive {MIMO} networks: A survey of opportunities,
  challenges and solutions,'' \emph{IEEE Communications Surveys \& Tutorials},
  vol.~24, no.~1, pp. 611--652, 2021.

\bibitem{elhoushy2021cell}
S.~Elhoushy, M.~Ibrahim, and W.~Hamouda, ``{Cell-free massive MIMO: A
  survey},'' \emph{IEEE Communications Surveys \& Tutorials}, vol.~24, no.~1,
  pp. 492--523, 2021.

\bibitem{liu2019spectral}
P.~Liu, K.~Luo, D.~Chen, and T.~Jiang, ``Spectral efficiency analysis of
  cell-free massive {MIMO} systems with zero-forcing detector,'' \emph{IEEE
  Transactions on Wireless Communications}, vol.~19, no.~2, pp. 795--807, 2019.

\bibitem{ngo2017cell}
H.~Q. Ngo, A.~Ashikhmin, H.~Yang, E.~G. Larsson, and T.~L. Marzetta,
  ``Cell-free massive {MIMO} versus small cells,'' \emph{IEEE Transactions on
  Wireless Communications}, vol.~16, no.~3, pp. 1834--1850, 2017.

\bibitem{chen2020structured}
S.~Chen, J.~Zhang, E.~Bj{\"o}rnson, J.~Zhang, and B.~Ai, ``Structured massive
  access for scalable cell-free massive {MIMO} systems,'' \emph{IEEE Journal on
  Selected Areas in Communications}, vol.~39, no.~4, pp. 1086--1100, 2020.

\bibitem{bjornson2020scalable}
E.~Bj{\"o}rnson and L.~Sanguinetti, ``Scalable cell-free massive {MIMO}
  systems,'' \emph{IEEE Transactions on Communications}, vol.~68, no.~7, pp.
  4247--4261, 2020.

\bibitem{CHEN2022695}
\BIBentryALTinterwordspacing
S.~Chen, J.~Zhang, J.~Zhang, E.~Björnson, and B.~Ai, ``A survey on
  user-centric cell-free massive {MIMO} systems,'' \emph{Digital Communications
  and Networks}, vol.~8, no.~5, pp. 695--719, 2022. [Online]. Available:
  \url{https://www.sciencedirect.com/science/article/pii/S2352864821001024}
\BIBentrySTDinterwordspacing

\bibitem{cao2023experimental}
Y.~Cao, P.~Wang, K.~Zheng, X.~Liang, D.~Liu, M.~Lou, J.~Jin, Q.~Wang, D.~Wang,
  Y.~Huang \emph{et~al.}, ``Experimental performance evaluation of cell-free
  massive {MIMO} systems using {COTS} {RRU} with {OTA} reciprocity calibration
  and phase synchronization,'' \emph{IEEE Journal on Selected Areas in
  Communications}, vol.~41, no.~6, pp. 1620--1634, 2023.

\bibitem{bjornson2019making}
E.~Bj{\"o}rnson and L.~Sanguinetti, ``Making cell-free massive {MIMO}
  competitive with {MMSE} processing and centralized implementation,''
  \emph{IEEE Transactions on Wireless Communications}, vol.~19, no.~1, pp.
  77--90, 2019.

\bibitem{wang2023full}
D.~Wang, X.~You, Y.~Huang, W.~Xu, J.~Li, P.~Zhu, Y.~Jiang, Y.~Cao, X.~Xia,
  Z.~Zhang \emph{et~al.}, ``Full-spectrum cell-free {RAN} for 6{G} systems:
  system design and experimental results,'' \emph{Science China Information
  Sciences}, vol.~66, no.~3, p. 130305, 2023.

\bibitem{heath2016overview}
R.~W. Heath, N.~Gonzalez-Prelcic, S.~Rangan, W.~Roh, and A.~M. Sayeed, ``An
  overview of signal processing techniques for millimeter wave {MIMO}
  systems,'' \emph{IEEE journal of selected topics in signal processing},
  vol.~10, no.~3, pp. 436--453, 2016.

\bibitem{alonzo2017cell}
M.~Alonzo and S.~Buzzi, ``Cell-free and user-centric massive {MIMO} at
  millimeter wave frequencies,'' in \emph{2017 IEEE 28th annual international
  symposium on personal, indoor, and mobile radio communications
  (PIMRC)}.\hskip 1em plus 0.5em minus 0.4em\relax IEEE, 2017, pp. 1--5.

\bibitem{alonzo2019energy}
M.~Alonzo, S.~Buzzi, A.~Zappone, and C.~D’Elia, ``Energy-efficient power
  control in cell-free and user-centric massive {MIMO} at millimeter wave,''
  \emph{IEEE Transactions on Green Communications and Networking}, vol.~3,
  no.~3, pp. 651--663, 2019.

\bibitem{femenias2019cell}
G.~Femenias and F.~Riera-Palou, ``Cell-free millimeter-wave massive {MIMO}
  systems with limited fronthaul capacity,'' \emph{IEEE Access}, vol.~7, pp.
  44\,596--44\,612, 2019.

\bibitem{meyer2022state}
E.~Meyer, D.~Kruglov, M.~Krivic, M.~Tanveer, R.~Argaez-Ramirez, Y.~Zhang, A.~B.
  Ojeda, K.~Smirnova, K.~Alekseev, M.~S. Mugisho \emph{et~al.}, ``The state of
  the art in beyond 5{G} distributed massive multiple-input multiple-output
  communication system solutions,'' \emph{Open Research Europe}, vol.~2, 2022.

\bibitem{yuan20223d}
Y.~Yuan, R.~He, B.~Ai, Z.~Ma, Y.~Miao, Y.~Niu, J.~Zhang, R.~Chen, and Z.~Zhong,
  ``A 3{D} geometry-based {TH}z channel model for 6{G} ultra massive {MIMO}
  systems,'' \emph{IEEE Transactions on Vehicular Technology}, vol.~71, no.~3,
  pp. 2251--2266, 2022.

\bibitem{xiao2017millimeter}
M.~Xiao, S.~Mumtaz, Y.~Huang, L.~Dai, Y.~Li, M.~Matthaiou, G.~K. Karagiannidis,
  E.~Bj{\"o}rnson, K.~Yang, I.~Chih-Lin \emph{et~al.}, ``Millimeter wave
  communications for future mobile networks,'' \emph{IEEE Journal on Selected
  Areas in Communications}, vol.~35, no.~9, pp. 1909--1935, 2017.

\bibitem{dai2022delay}
L.~Dai, J.~Tan, Z.~Chen, and H.~V. Poor, ``Delay-phase precoding for wideband
  {TH}z massive {MIMO},'' \emph{IEEE Transactions on Wireless Communications},
  vol.~21, no.~9, pp. 7271--7286, 2022.

\bibitem{rappaport2013millimeter}
T.~S. Rappaport, S.~Sun, R.~Mayzus, H.~Zhao, Y.~Azar, K.~Wang, G.~N. Wong,
  J.~K. Schulz, M.~Samimi, and F.~Gutierrez, ``Millimeter wave mobile
  communications for 5{G} cellular: It will work!'' \emph{IEEE access}, vol.~1,
  pp. 335--349, 2013.

\bibitem{liu2014phase}
A.~Liu and V.~Lau, ``Phase only {RF} precoding for massive {MIMO} systems with
  limited {RF} chains,'' \emph{IEEE Transactions on Signal Processing},
  vol.~62, no.~17, pp. 4505--4515, 2014.

\bibitem{nguyen2022hybrid}
N.~T. Nguyen, K.~Lee, and H.~Dai, ``Hybrid beamforming and adaptive {RF} chain
  activation for uplink cell-free millimeter-wave massive {MIMO} systems,''
  \emph{IEEE Transactions on Vehicular Technology}, vol.~71, no.~8, pp.
  8739--8755, 2022.

\bibitem{gao2016energy}
X.~Gao, L.~Dai, S.~Han, I.~Chih-Lin, and R.~W. Heath, ``Energy-efficient hybrid
  analog and digital precoding for mm{W}ave {MIMO} systems with large antenna
  arrays,'' \emph{IEEE Journal on Selected Areas in Communications}, vol.~34,
  no.~4, pp. 998--1009, 2016.

\bibitem{sohrabi2016hybrid}
F.~Sohrabi and W.~Yu, ``Hybrid digital and analog beamforming design for
  large-scale antenna arrays,'' \emph{IEEE Journal of Selected Topics in Signal
  Processing}, vol.~10, no.~3, pp. 501--513, 2016.

\bibitem{brady2013beamspace}
J.~Brady, N.~Behdad, and A.~M. Sayeed, ``Beamspace {MIMO} for millimeter-wave
  communications: System architecture, modeling, analysis, and measurements,''
  \emph{IEEE Transactions on Antennas and Propagation}, vol.~61, no.~7, pp.
  3814--3827, 2013.

\bibitem{amadori2015low}
P.~V. Amadori and C.~Masouros, ``Low {RF}-complexity millimeter-wave
  beamspace-{MIMO} systems by beam selection,'' \emph{IEEE Transactions on
  Communications}, vol.~63, no.~6, pp. 2212--2223, 2015.

\bibitem{guo2018joint}
R.~Guo, Y.~Cai, M.~Zhao, Q.~Shi, B.~Champagne, and L.~Hanzo, ``Joint design of
  beam selection and precoding matrices for mm{W}ave {MU}-{MIMO} systems
  relying on lens antenna arrays,'' \emph{IEEE Journal of Selected Topics in
  Signal Processing}, vol.~12, no.~2, pp. 313--325, 2018.

\bibitem{zeng2016millimeter}
Y.~Zeng and R.~Zhang, ``Millimeter wave {MIMO} with lens antenna array: A new
  path division multiplexing paradigm,'' \emph{IEEE Transactions on
  Communications}, vol.~64, no.~4, pp. 1557--1571, 2016.

\bibitem{zhu2009chunk}
H.~Zhu and J.~Wang, ``{Chunk-based resource allocation in OFDMA systems-part I:
  chunk allocation},'' \emph{IEEE Transactions on Communications}, vol.~57,
  no.~9, pp. 2734--2744, 2009.

\bibitem{zhu2011chunk}
H.~\vspace{0mm}Zhu and J.~Wang, ``{Chunk-based resource allocation in OFDMA
  systems—Part II: Joint chunk, power and bit allocation},'' \emph{IEEE
  Transactions on Communications}, vol.~60, no.~2, pp. 499--509, 2011.

\bibitem{yan2019asynchronous}
H.~Yan and I.-T. Lu, ``Asynchronous reception effects on distributed massive
  {MIMO}-{OFDM} system,'' \emph{IEEE Transactions on Communications}, vol.~67,
  no.~7, pp. 4782--4794, 2019.

\bibitem{li2021impacts}
J.~Li, M.~Liu, P.~Zhu, D.~Wang, and X.~You, ``Impacts of asynchronous reception
  on cell-free distributed massive {MIMO} systems,'' \emph{IEEE Transactions on
  Vehicular Technology}, vol.~70, no.~10, pp. 11\,106--11\,110, 2021.

\bibitem{li2023cell}
G.~Li, S.~Wu, C.~You, W.~Zhang, G.~Shang, and X.~Zhou, ``Cell-free massive
  {MIMO}-{OFDM}: Asynchronous reception and performance analysis,'' \emph{IEEE
  Internet of Things Journal}, 2023.

\bibitem{zheng2023asynchronous}
J.~Zheng, J.~Zhang, J.~Cheng, V.~C. Leung, D.~W.~K. Ng, and B.~Ai,
  ``Asynchronous cell-free massive {MIMO} with rate-splitting,'' \emph{IEEE
  Journal on Selected Areas in Communications}, vol.~41, no.~5, pp. 1366--1382,
  2023.

\bibitem{zheng2022performance}
J.~Zheng, Z.~Zhao, J.~Zhang, J.~Cheng, and V.~C. Leung, ``Performance analysis
  of cell-free massive {MIMO} systems with asynchronous reception,'' in
  \emph{2022 IEEE Globecom Workshops (GC Wkshps)}.\hskip 1em plus 0.5em minus
  0.4em\relax IEEE, 2022, pp. 190--195.

\bibitem{jafri2024cooperative}
M.~Jafri, S.~Srivastava, P.~Kumar, A.~K. Jagannatham, and L.~Hanzo,
  ``Cooperative hybrid beamforming for the mitigation of realistic asynchronous
  interference in cell-free mm{W}ave {MIMO} networks,'' \emph{IEEE Transactions
  on Communications}, 2024.

\bibitem{li2024cell}
G.~Li, C.~You, S.~Wu, W.~Zhang, G.~Shang, and X.~Zhou, ``Cell-free massive
  {MIMO}-{OFDM}: Asynchronous reception and uplink performance analysis,'' in
  \emph{2024 IEEE Wireless Communications and Networking Conference
  (WCNC)}.\hskip 1em plus 0.5em minus 0.4em\relax IEEE, 2024, pp. 1--6.

\bibitem{you2017bdma}
L.~You, X.~Gao, G.~Y. Li, X.-G. Xia, and N.~Ma, ``{BDMA} for
  millimeter-wave/terahertz massive {MIMO} transmission with per-beam
  synchronization,'' \emph{IEEE Journal on Selected Areas in Communications},
  vol.~35, no.~7, pp. 1550--1563, 2017.

\bibitem{gao2016fast}
X.~Gao, L.~Dai, Y.~Zhang, T.~Xie, X.~Dai, and Z.~Wang, ``Fast channel tracking
  for terahertz beamspace massive {MIMO} systems,'' \emph{IEEE Transactions on
  Vehicular Technology}, vol.~66, no.~7, pp. 5689--5696, 2016.

\bibitem{gao2019wideband}
X.~Gao, L.~Dai, S.~Zhou, A.~M. Sayeed, and L.~Hanzo, ``Wideband beamspace
  channel estimation for millimeter-wave {MIMO} systems relying on lens antenna
  arrays,'' \emph{IEEE Transactions on Signal Processing}, vol.~67, no.~18, pp.
  4809--4824, 2019.

\bibitem{chowdhury2023resilient}
A.~Chowdhury and C.~R. Murthy, ``How resilient are cell-free massive {MIMO}
  {OFDM} systems to propagation delays?'' in \emph{2023 IEEE 24th International
  Workshop on Signal Processing Advances in Wireless Communications
  (SPAWC)}.\hskip 1em plus 0.5em minus 0.4em\relax IEEE, 2023, pp. 581--585.

\bibitem{yang2018pilot}
F.~Yang, P.~Cai, H.~Qian, and X.~Luo, ``Pilot contamination in massive {MIMO}
  induced by timing and frequency errors,'' \emph{IEEE Transactions on Wireless
  Communications}, vol.~17, no.~7, pp. 4477--4492, 2018.

\bibitem{sun2019beam}
X.~Sun, C.~Qi, and G.~Y. Li, ``Beam training and allocation for multiuser
  millimeter wave massive {MIMO} systems,'' \emph{IEEE Transactions on Wireless
  Communications}, vol.~18, no.~2, pp. 1041--1053, 2019.

\end{thebibliography}

\end{document}